\documentclass{CSML}
\pdfoutput=1

\usepackage{lastpage}
\newcommand{\dOi}{14(1:1)2018}
\lmcsheading{}{1--\pageref{LastPage}}{}{}%
{Nov.~14,~2016}{Jan.~09,~2018}{}


\ACMCCS{[{\bf Theory of computation}]: Concurrency}

\usepackage{hyperref}
\hypersetup{hidelinks}

\usepackage[utf8]{inputenc}
\usepackage[english]{babel}

\usepackage{amsmath,amssymb,mathrsfs,stmaryrd,xspace}
\usepackage{graphicx,url,color,eurosym,wrapfig,alltt}
\usepackage{tabularx,multicol,multirow,subfigure}
\usepackage{comment,rotating}

\usepackage{tikz}
\usetikzlibrary{arrows,automata,shadows}
\usetikzlibrary{shapes,calc}

\makeatletter
\newcommand{\xleftrightarrow}[2][]{\ext@arrow 3359\leftrightarrowfill@{#1}{#2}}
\makeatother

\newcommand{\xdasharrow}[2][->]{
\tikz[baseline=-\the\dimexpr\fontdimen22\textfont2\relax]{
\node[anchor=south,font=\scriptsize, inner ysep=1.5pt,outer xsep=2.2pt](x){#2};
\draw[shorten <=3.4pt,shorten >=3.4pt,dashed,#1](x.south west)--(x.south east);
}
}

\newcommand{\syc}{\xleftrightarrow{\text{\,\,\,\,\,\,}}}
\newcommand{\wsc}{\xdasharrow[<->]{\text{\,\,\,\,\,\,\,}}}

\newcommand{\ac}{\xleftrightarrow{\text{\,\,a\,\,}}}
\newcommand{\wac}{\xdasharrow[<->]{\text{\,\,a\,\,}}}

\newcommand{\stimes}{\otimes}
\newcommand{\atimes}{\otimes_{\mathit{as}}}

\newcommand{\renBA}{B_{\outBA}^\rhd}
\newcommand{\renAB}{A_{\outAB}^\rhd}

\newcommand{\reachBA}{\reach(A \stimes \renBA)}
\newcommand{\reachAB}{\reach(\renAB \stimes B)}

\newcommand{\df}{\textit{df}(A \stimes B)}
\newcommand{\dfA}{\textit{df}(A \atimes B)}

\newcommand{\emptyseq}{\epsilon}

\newcommand{\reach}[1]{\mathcal{R}#1}
\newcommand{\xmust}[3]{{\xrightarrow{\,#1\,}}\!\!\!{\genfrac{}{}{0pt}{}{#2}{#3}}}
\newcommand{\Must}[3]{{\stackrel{#1}{\Longrightarrow}\!\!\!{\genfrac{}{}{0pt}{}{#2}{#3}}}}

\newcommand{\outAB}{\out_{\mathit{AB}}}
\newcommand{\outBA}{\out_{\mathit{BA}}}

\newcommand{\xtRarrA}[1] { % triple rightarrow
  \stackrel{#1}{\Rrightarrow}_{_{\mkern-8.mu\mathit{A}}}
}
\newcommand{\xtRarrB}[1] { % triple rightarrow
  \stackrel{#1}{\Rrightarrow}_{_{\mkern-8.mu\mathit{B}}}
}

\newcommand{\states}{\mathit{states}}
\newcommand{\start}{\mathit{start}}
\newcommand{\act}{\mathit{act}}

\newcommand{\inp}{\mathit{in}}
\newcommand{\out}{\mathit{out}}
\newcommand{\internal}{\mathit{int}}
\newcommand{\must}[3]{{\stackrel{#1}{\longrightarrow}\!\!\!{\genfrac{}{}{0pt}{}{#2}{#3}}}}

\newcommand{\shared}{\mathit{shared}}

\begin{document}

\title[COMPATIBILITY PROPERTIES OF COMMUNICATING COMPONENTS]{Compatibility Properties of Synchronously and Asynchronously Communicating Components}
\author{Rolf Hennicker}
\address{Ludwig-Maximilians-Universität München, Germany}
\email{hennicker@ifi.lmu.de}

\author{ Michel Bidoit}
\address{LSV, CNRS and ENS de Cachan, France}
\email{bidoit@lsv.ens-cachan.fr}

\begin{abstract}
We study interacting components and their compatibility with respect to synchronous and asynchronous composition. The behavior of components is formalized by I/O-transition systems. Synchronous composition is based on simultaneous execution of shared output and input actions of two components while asynchronous composition uses unbounded FIFO-buffers for message transfer. In both contexts we study compatibility notions based on the idea that any output issued by one component should be accepted as an input by the other. We distinguish between strong and weak versions of compatibility, the latter allowing the execution of internal actions before a message is accepted. We consider open systems and study conditions under which (strong/weak) \emph{synchronous} compatibility is sufficient and necessary to get (strong/weak) \emph{asynchronous} compatibility. We show that these conditions characterize half-duplex systems. Then we focus on the verification of weak asynchronous compatibility for possibly non half-duplex systems and provide a decidable criterion that ensures weak asynchronous compatibility. We investigate conditions under which this criterion is complete, i.e. if it is not satisfied then the asynchronous system is not weakly asynchronously compatible. Finally, we discuss deadlock-freeness and investigate relationships between deadlock-freeness in the synchronous and in the asynchronous case. 
\end{abstract}

\maketitle

\section{Introduction}\label{sec:intro}

Distributed systems consist of sets of components which are deployed on different nodes and communicate through certain media.
In this work we consider reactive components with a well defined behavior which communicate by message exchange.
Each single component has a life cycle during which it sends and receives messages and it can also perform internal actions
in between. For the correct functioning of the overall system it is essential that no communication errors occur
during component interactions.
Two prominent classes of communication errors can be distinguished:
The first one concerns situations, in which an output of one component is not accepted as an input by the other.
The second one occurs if a component waits for an input which is never delivered.
In this paper we focus on the former kind of communication error and we
consider systems consisting of two components.
We assume that outputs are autonomous actions and we call two components compatible
if any output issued by one component is accepted by its communication partner.
In our study
we deal with bidirectional, peer to peer communication and with synchronous and asynchronous
message exchange. The former is based on a rendezvous mechanism such that two components
must execute shared output and input actions together while the latter uses potentially unbounded FIFO-buffers which hold
the messages sent by one component and received by the other.
We consider FIFO-buffers since these are used in well-known communication models, like CFSMs (Communicating Finite State Machines~\cite{Brand-Zafiropulo}), but also many concrete technologies rely on FIFO-communication, like the TCP protocol,  the Java Messaging Service (being part of the Java Enterprise Edition
as a message oriented middleware) and the Microsoft Message Queuing Service for service-oriented architectures. While compatibility of synchronously communicating components is decidable, see, e.g.,~\cite{Alfaro2001},
it is undecidable if unbounded FIFO-buffers are used as communication channels~\cite{Brand-Zafiropulo}. Therefore we are interested to investigate effective proof techniques for the verification of compatibility in the asynchronous case.

For this purpose we study, in the first part of this paper, relationships between synchronous and asynchronous compatibility. 
In each case we consider two versions, a strong and a weak compatibility notion.
For the formalization of component behaviors we use I/O-transition systems (IOTSes) and call
two IOTses $A$ and $B$ \emph{strongly synchronously compatible} if in any reachable state of the
synchronous product of $A$ and $B$, if one component, say $A$, has a transition enabled with
output action $a$ then $B$ must have a transition enabled with input action $a$.
In many practical examples it turns out that before interacting with the sending component
the receiving component should still be able to perform some internal actions in between. 
This leads to our notion of \emph{weak synchronous compatibilty}.
In the asynchronous context, components communicate via unbounded message queues. The idea of asynchronous compatibility is to require
that whenever a message queue is not empty, then the receiver component must be able to take the next
element of the queue; a property called specified reception in~\cite{Brand-Zafiropulo}. We distinguish again between \emph{strong} and \emph{weak} versions of \emph{asynchronous compatibility}.
In the asynchronous context the weak compatibility notion is particularly powerful since it allows a component,
before it inputs a message waiting in the queue, still to put itself messages in its output queue (since we consider such enqueue actions as internal). 

An obvious question
is to what extent synchronous and asynchronous compatibility notions can be related to each other and, if this is not possible,
which proof techniques can be used to verify asynchronous compatibility. 
We contribute to these issues with the following results:

\begin{enumerate}
\item
We establish a relationship between strong/weak synchronous and asynchronous compatibility of two components (Sects.~\ref{sec:sync2asynch} and~\ref{sec:async2synch}). % for the strong and for the weak case.
As a main result (Cor.~\ref{cor:syncIFFasync}) we get that strong (weak) synchronous compatibility is equivalent to strong (weak) asynchronous compatibility if the system enjoys the \emph{half-duplex} property~\cite{DBLP:journals/iandc/CeceF05}.
This means that in the asynchronous system at any time at most one message queue is not empty.
%After defining the compatibility notions (Sect.\ref{sec:comp}),
%we establish a relationship between strong/weak synchronous and strong/weak asynchronous compatibility of two communicating components (Sects.~\ref{sec:sync2asynch} and~\ref{sec:async2synch}). % for the strong and for the weak case.
%For this purpose we formulate three equivalent (and decidable) conditions which guarantee that strong/weak synchronous compatibility is necessary and sufficient for strong/weak asynchronous compatibility. One of the three conditions requires that the asynchronous system is half-duplex, i.e.\ at any time at most one of the two message queues is not empty; see~\cite{DBLP:journals/iandc/CeceF05}.

\item
In the second part of this work (Sect.~\ref{sec:asynch-comp}), we consider general, possibly non half-duplex systems and study the verification of weak asynchronous compatibility in such cases. 
In Sect.~\ref{sec:criterion} we investigate a decidable and powerful criterion, called WAC-criterion, for weak asynchronous compatibility (Thm.~\ref{thm:general-case}).
The criterion cannot be necessary, since for unbounded FIFO-buffers the problem is undecidable, i.e., our proof method cannot be complete. In Sect.~\ref{sec:completeness} we discuss
how far we are away from completeness. To this end we develop a decidable
completeness criterion (Thm.~\ref{thm.completenss}). If this criterion is satisfied
then we can even disprove weak asynchronous compatibility. 
\end{enumerate}

\begin{figure}
  \centering
 \includegraphics[scale=0.5]{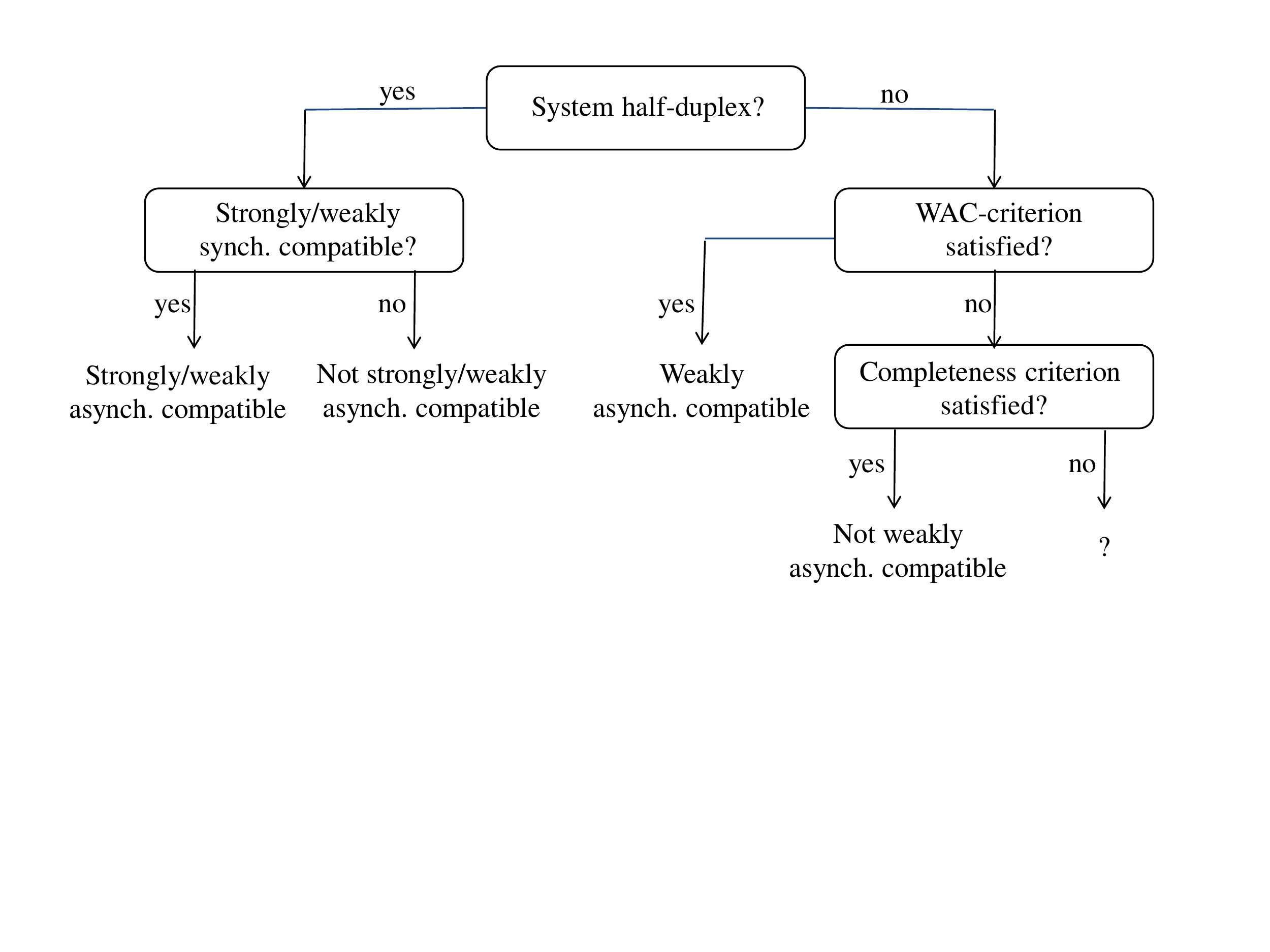}
\vspace{-35mm}\caption{Verification method for compatibility}
  \label{fig:verification-method}
\end{figure}

\noindent
Our results lead to the verification methodology for compatibility checking summarized in
Fig.~\ref{fig:verification-method}.
Assume given two asynchronously communicating components $A$ and $B$, each one having finitely many local states.
First,  we check whether the system is half-duplex, which is decidable~\cite{DBLP:journals/iandc/CeceF05}.
If the answer is positive, then we check whether $A$ and $B$ are 
strongly (weakly) synchronously compatible, which is decidable as well~\cite{tacas2010}.
Then the equivalence in Cor.~\ref{cor:syncIFFasync} shows that
if the answer is positive, then $A$ and $B$ are 
strongly (weakly) asynchronously compatible, otherwise they are not.
If the system is not half-duplex we proceed as follows:
We check whether the WAC-criterion (Thm.~\ref{thm:general-case})
is satisfied, which is decidable.
If the answer is positive, then $A$ and $B$ are 
weakly asynchronously compatible.
Otherwise we check the completeness criterion (Thm.~\ref{thm.completenss}),
which is decidable as well. If the answer is positive, we know that
$A$ and $B$ are not weakly asynchronously compatible.
Otherwise we don't know.

In Sect.~\ref{sec:df}, we discuss deadlock-freeness of synchronous and asynchronous systems and show that deadlock-freeness is neither sufficient nor necessary for compatibility. 
We show how to perform a deadlock analysis for asynchronous systems following again the idea to prove properties of the synchronous system in order to get properties of the asynchronous one.
Sect.~\ref{sec:related-work} discusses related work and
Sect.~\ref{sec:conclusion} summarizes our results and future work.%discusses next steps.

This paper is a significantly revised and extended version of the conference paper~\cite{DBLP:conf/coordination/HennickerBD16}.
In Sect.~\ref{sec:asynch-comp} we have simplified the criterion for weak asynchronous compatibility. We have also added Sect.~\ref{sec:completeness}, which discusses completeness of the criterion.
Moreover, the analysis of weak asynchronous compatibility is complemented in the new Sect.~\ref{sec:df}  by
a deadlock analysis.

%This paper is a revised and extended version of the conference paper~\cite{DBLP:conf/coordination/HennickerBD16}.
%In Sect.~\ref{sec:asynch-comp} we have simplified the criterion for weak asynchronous compatibility. We have also added the subsection~\ref{sec:completeness}, which discusses completeness of the criterion.
%Moreover, the analysis of weak asynchronous compatibility is complemented in the new Sect.~\ref{sec:df}  by
%a deadlock analysis for asynchronous systems following again the idea to prove properties for the synchronous system in order to get properties of the asynchronous one.
%For all results detailed proofs are provided in this paper. % which were not included in~\cite{DBLP:conf/coordination/HennickerBD16}.

%\ref{sec:asynch-comp}
%\ref{sec:criterion}
%\ref{sec:completeness}

%%%%%%%%%%%%%%%%%%%%%%%%%%%%%%%%%%%%%%%%%%%%%%%%%%%%%%%
%

\section{I/O-Transition Systems and Their Compositions}\label{sec:iolts}

We start with the definitions of I/O-transition systems and
their synchronous and asynchronous compositions which are the basis of the subsequent study.
%\begin{align*}
%&\text{(action prefix)} &&a . P \rightarrow{a} P\\[4mm]
%&\text{(choice-left)} && \frac{P \rightarrow{a} P'}{P+Q \rightarrow{a} P'}\\[4mm]
%&\text{(choice-right)} && \frac{Q \rightarrow{a} Q'}{P+Q \rightarrow{a} Q'}\\[4mm]
%&\text{(type declaration)} && \frac{P \rightarrow{a} P'}{PN \rightarrow{a} P'}~ if~ PN \stackrel{\text{def}}{=} P
%\end{align*}

\begin{defi}[IOTS]
  An \emph{I/O-transition system} is a quadruple
$A = (\states_A,\start_A,$ $ \act_A, \must{}{}{A})$ consisting of a set of states $\states_A$, an initial state $\start_A \in
\states_A$, a set $\act_A = \inp_A \cup \out_A \cup \internal_A$ of actions being the disjoint union of
sets $\inp_A$, $\out_A$ and $\internal_A$ of input, output and internal actions resp., and a
transition relation $\must{}{}{A} \subseteq \states_A \times act_A \times \states_A$.
%The action set $act_A$ and its partition into input, output and internal actions is called \emph{labeling} of $A$.
\end{defi}

We write $s \must{a}{}{A} s'$ instead of $(s,a,s') \in  \must{}{}{A}$.
%If we want to emphasize that $a$ is an input action (ouput action) of $A$
%we write $s \must{a?}{}{A} s'$ ($s \must{a!}{}{A} s'$ resp.).
%If we want to emphasize that $a$ is an internal action of $A$,
%we write $s \must{\tau_a}{}{A} s'$.
For $X \subseteq \act_A$
we write $s \xmust{X}{*}{A} s'$ if there exists a (possibly empty) sequence of transitions
$s \must{a_1}{}{A} s_1 \ldots s_{n-1} \must{a_n}{}{A} s'$
%$s \must{a_1}{}{A} s_1 \must{a_2}{}{A} \ldots s_{n-1} \must{a_n}{}{A} s'$
involving only actions of $X$, i.e. $a_1,\ldots,a_n \in X$.
A state $s \in \states_A$ is \emph{reachable} if $\start_A \xmust{\act_A}{*}{A} s$.
The set of reachable states of $A$ is denoted by  $\reach(A)$.
%The class of I/O-transition systems is denoted by $\mathcal{T}$.

%\paragraph{Composability and Synchronous Composition.}
Two IOTSes $A$ and $B$
are \emph{(syntactically) composable} if their actions only overlap on
complementary types, i.e.  $\act_A \cap \act_B = (in_A \cap
out_B) \cup (in_B \cap out_A)$. The \emph{set of shared actions} $\act_A \cap
\act_B$ is denoted by $\shared(A,B)$.  The \emph{synchronous
  composition} of two IOTSes $A$ and $B$ is
defined as the product of transition systems with
synchronization on shared actions which become internal actions in the
composition. Shared actions can only be executed together; they are blocked
if the other component is not ready for communication. 
%cannot be executed freely by one component only.
%In contrast internal actions are always performed by a single component only and the same holds
%for non-shared input and output actions which are open for the environment of the composition.
In contrast, internal actions and non-shared input and output actions %which are open for the environment of the composition
can always be executed by a single component in the composition. 
These (non-shared) actions are called \emph{free actions} in the following.

\begin{defi}[Synchronous composition]
  Let $A$ and $B$ be two composable IOTSes. The
  \emph{synchronous composition of $A$ and $B$} is the IOTS
$A \stimes B = (\states_A \times \states_B, (\start_A,\start_B),$  $\act_{A \stimes B},  \must{}{}{A \stimes B})$
where $\act_{A \stimes B}$ is the
disjoint union of the input actions $\inp_{A \stimes B} = (\inp_A \cup in_B) \smallsetminus
  \shared(A,B)$, the output actions $\out_{A \stimes B} = (\out_A \cup out_B) \smallsetminus
  \shared(A,B)$, and the internal actions $\internal_{A \stimes B} = \internal_A \cup \internal_B \cup
  \shared(A,B)$. The transition relation of $A \stimes B$ is the
  smallest relation such that
\begin{itemize}
\item for all $a \in \act_A\smallsetminus\shared(A,B)$,
    if $s \must{a}{}{A} s'$,
    then $(s,t) \must{a}{}{A \stimes B} (s',t)$ for all $t \in \states_B$,
\item for all $a \in \act_B\smallsetminus\shared(A,B)$,
if $t \must{a}{}{B} t'$,
    then $(s,t) \must{a}{}{A \stimes B} (s,t')$ for all $s \in \states_A$, and
\item for all $a \in \shared(A,B)$,
   if $s \must{a}{}{A} s'$ and $t \must{a}{}{B} t'$,
    then $(s,t) \must{a}{}{A \stimes B} (s',t')$.
\end{itemize}
\end{defi}

The synchronous composition of two IOTSes $A$ and $B$ yields a \emph{closed} system
if it has no input and output actions, i.e.\
$(\inp_A \cup in_B) \smallsetminus \shared(A,B) = \emptyset$ and $(\out_A \cup out_B) \smallsetminus \shared(A,B) = \emptyset$,
otherwise the system is \emph{open}.

\begin{figure}
  \centering
 \includegraphics[scale=0.4]{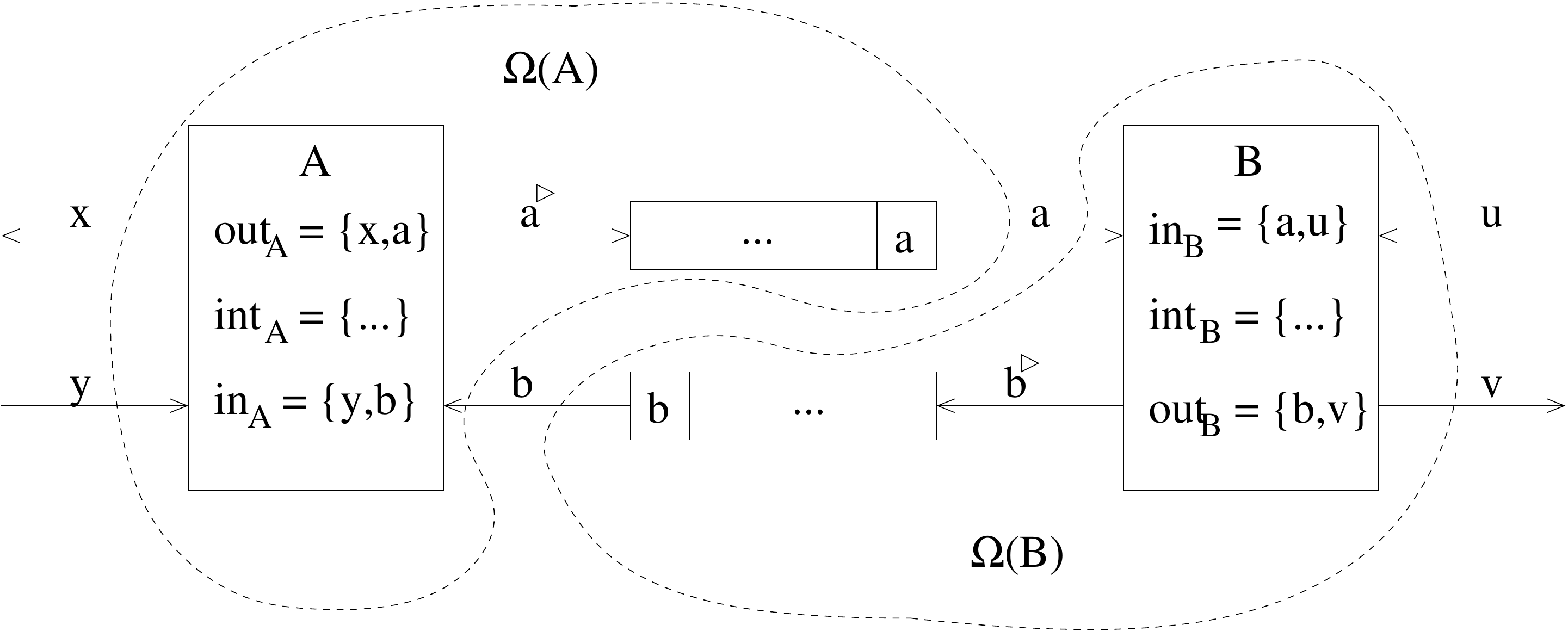}
\caption{Asynchronous communication}
  \label{fig:asynchcomposition}
\end{figure}

In distributed applications, implemented, e.g., with a message-passing
middleware, usually an asynchronous communication pattern is used. In this paper,
we consider asynchronous communication via unbounded message queues. 
In Fig.~\ref{fig:asynchcomposition} two asynchronously communicating IOTSes $A$ and $B$
are depicted.
$A$ sends a message $a$ to $B$ by putting it, with action $a^\rhd$, into a queue which stores
the outputs of $A$.
Then $B$ can receive $a$ by removing it, with action $a$, from
the queue. In contrast to synchronous communication, the sending of a message
cannot be blocked if the receiver is not ready to accept it.
Similarly, $B$ can send a message $b$ to $A$ by using a second queue which
stores the outputs of $B$.
The system in Fig~\ref{fig:asynchcomposition} is open: $A$ has an open output $x$
to the environment and an open input $y$ for messages coming from the environment.
Similarly $B$ has an open input $u$ and an open output $v$.
Additionally, $A$ and $B$ may have some internal actions. 

To formalize asynchronous communication, we equip each communicating IOTS with an
``output queue'', which leads to a new IOTS indicated in Fig.~\ref{fig:asynchcomposition}
by $\Omega(A)$ and $\Omega(B)$ respectively. 
The motivation for using output queues should become clear when we define asynchronous compatibility in the next section.
Formally, we represent an output queue as an (infinite) IOTS
and then, in the case of $A$, we compose it with a renamed version of $A$ where all outputs $a$ of $A$
(to be stored in the queue) are renamed to enqueue actions of the form $a^\rhd$. 

\begin{defi}[IOTS with output queue]\label{def:output-queue}\hfill
\begin{enumerate}
\item Let $M$ be a set of names and $M^\rhd = \{ a^\rhd \mid a \in M\}$.
The \emph{queue IOTS for $M$} is
$Q_M = (M^*, \emptyseq, \act_{Q_M}, \must{}{}{Q_M})$ where
  the set of states is the set $M^*$ of all words over
  $M$, the initial state $\emptyseq \in M^*$ is the empty word, and
  the set of actions $\act_{Q_M}$ is the disjoint union of
  input actions $\inp_{Q_M} = M^\rhd$, output actions $\out_{Q_M} = M$ and with no internal action.
%The actions in  $M^\rhd$ are called ``enqueue actions''.
The transition relation $\must{}{}{Q_M}$ is the smallest relation such that
  \begin{itemize}
  \item for all $a^\rhd \in M^\rhd$ and states $q \in M^*: \ q \, \must{a^\rhd}{}{Q_M}\, qa$ (enqueue on the right),
  \item for all $a \in M$ and states $q \in M^*: \ aq \, \must{a}{}{Q_M}\, q$ (dequeue on the left).
  \end{itemize}

%Let $A$ be an IOTS with actions $\act_A$ and $M \subseteq \out_A$ such that $M^\rhd \cap \act_A = \emptyset$.
%\vspace{2mm}
\item
 Let $A$ be an IOTS such that $M \subseteq \out_A$ and $M^\rhd \cap \act_A = \emptyset$.
Let $A_{M}^\rhd$ be the renamed version of $A$ where all $a \in M$ are renamed to $a^\rhd$.
%be a subset of the output actions of $A$.
The \emph{IOTS $A$ equipped with output queue for $M$} is given by
  the synchronous composition $\Omega_{M}(A) = A_{M}^\rhd \stimes Q_M$. (Note that $A_{M}^\rhd$ and $Q_M$ are composable.)
%where $A_{M}^\rhd$ denotes the renamed version of $A$ where all $a \in M$ are renamed to $a^\rhd$.
%Obviously, $A_{M}^\rhd$ and $Q_M$ are composable.
% and therefore the product $\Omega_{M}(A)$ is well defined.
\end{enumerate}
\end{defi}

The states of $\Omega_{M}(A)$ are pairs $(s,q)$ where $s$ is a state of $A$ and $q$ is a word over $M$.
The initial state is $(\start_A,\emptyseq)$. For the actions we have $\inp_{\Omega_{M}(A)} = \inp_A, \out_{\Omega_{M}(A)} = \out_A$,
and $\internal_{\Omega_{M}(A)} = \internal_A \cup M^\rhd$.
%The input and the output actions of $\Omega_{M}(A)$ coincide
%with the ones of $A$.
%The internal actions of $\Omega_{M}(A)$ are either internal actions of $A$ or, for $a \in M$,
%enqueue actions $a^\rhd$. 
Transitions in $\Omega_{M}(A)$ are:

\begin{itemize}
\item
if $a \in \inp_A$ and $s  \must{a}{}{A} s'$
then $(s,q) \must{a}{}{\Omega_{M}(A)} (s',q)$,%\\
%(input of $A$  and hence of $\Omega_{M}(A)$), 
\item
if $a \in \out_A \smallsetminus M$ and $s  \must{a}{}{A} s'$
then $(s,q) \must{a}{}{\Omega_{M}(A)} (s',q)$,%\\
%(output of $A$ and hence of  $\Omega_{M}(A)$ for actions not belonging to $M$),
\item
if $a \in M \subseteq \out_A$
then $(s,a q) \must{a}{}{\Omega_{M}(A)} (s,q) $,%\\
%(output of  $Q_M$ and hence of $\Omega_{M}(A)$ removing $a$ from the queue),
\item
if $a \in \internal_A$ and $s  \must{a}{}{A} s'$
then $(s,q) \must{a}{}{\Omega_{M}(A)} (s',q)$,%\\
%(internal action of $A$ and hence of  $\Omega_{M}(A)$),
\item
 if $a^\rhd \in  M^\rhd$ and $s  \must{a}{}{A} s'$ (i.e.\, $s  \must{a^\rhd}{}{A_{M}^\rhd} s'$)
then $(s,q) \must{a^\rhd}{}{\Omega_{M}(A)} (s',q a) $.%\\
%(internal action of $\Omega_{M}(A)$ putting $a$ in the queue by synchronizing output $a^\rhd$ of $A_{M}^\rhd$ and input $a^\rhd$ of $Q_M$).
\end{itemize}

To define the asynchronous composition of two IOTSes $A$ and $B$, we assume that $A$ and $B$
 are \emph{asynchronously composable} which means that $A$ and $B$ are composable (as before) and
$\shared(A,B)^\rhd \cap (\act_A \cup \act_B) = \emptyset$, i.e.\ no name conflict can arise when we
rename a shared action $a$ to $a^\rhd$. Concerning $A$ we consider the output actions of $A$ which are shared with
input actions of $B$ and denote them by $\outAB = \out_A \cap \inp_B$. These are the messages
of $A$ directed to $B$. Then, according to Def.~\ref{def:output-queue},
the IOTS $A$ equipped with output queue for $\outAB$ is given by
$\Omega_{\outAB}(A) = A_{\outAB}^\rhd \stimes Q_{\outAB}$.
Note that $ A_{\outAB}^\rhd$ is the renamed version of $A$ where all actions $a \in \outAB$ are renamed to $a^\rhd$.  
Similarly,  we consider the output actions of $B$ which are shared with input actions of $A$, denote them by $\outBA = \out_B \cap \inp_A$ and construct the IOTS $\Omega_{\outBA}(B)  = B_{\outBA}^\rhd \stimes Q_{\outBA}$ which represents the component $B$ equipped with output queue for $\outBA$.
The IOTSes  $\Omega_{\outAB}(A)$ and $\Omega_{\outBA}(B)$ are then synchronously composed which
gives the asynchronous composition of $A$ and $B$.

\begin{defi}[Asynchronous composition]\label{def:asynch-composition}
 Let $A$, $B$  be two asynchronously composable IOTSes.
  The \emph{asynchronous composition} of $A$ and $B$ is defined by
$A \atimes B = \Omega_{\outAB}(A) \stimes \Omega_{\outBA}(B)$.\footnote{Note that 
$\Omega_{\outAB}(A)$ and $\Omega_{\outBA}(B)$ are composable.}
\end{defi}

In the sequel we will briefly write $\Omega(A)$ for $\Omega_{\outAB}(A)$ and $\Omega(B)$ for $\Omega_{\outBA}(B)$.
%In the sequel we will write $\Omega(A)$ for $\Omega_{M_A}(A)$ and $\Omega(B)$ for $\Omega_{M_B}(B)$.
The states of $\Omega(A) \stimes \Omega(B)$ are pairs $((s_A,q_A),(s_B,q_B))$ where $s_A$ is a state of $A$, the queue $q_A$
stores elements of $\outAB$, $s_B$ is a state of $B$, and the queue $q_B$ stores elements of $\outBA$.
The initial state is $((\start_A,\emptyseq),(\start_B,\emptyseq))$.
For the actions we have
$\inp_{\Omega(A) \stimes \Omega(B)} = \inp_{A \stimes B}$,
$\out_{\Omega(A) \stimes \Omega(B)} = \out_{A \stimes B}$,
and $\internal_{\Omega(A) \stimes \Omega(B)} = \internal_{A \stimes B} \cup \shared(A,B)^\rhd$.
For the transitions in $\Omega(A) \stimes \Omega(B)$ we have two main cases:

\begin{enumerate}
\item
Transitions which can freely occur in $A$ or in $B$ without involving any output queue. These transitions change just the local state of $A$ or of $B$. An example would be
a transition $s_A  \must{a}{}{A} s'_A$ with action $a \in \out_A \smallsetminus \inp_B$ which induces a transition
$((s_A,q_A),(s_B,q_B))$  $\must{a}{}{\Omega(A) \stimes \Omega(B)}$  $((s'_A,q_A),(s_B,q_B))$.

\item
Transitions which involve the output queue of $A$.
There are two sub-cases concerning dequeue and enqueue actions which are internal actions in
$\Omega(A) \stimes \Omega(B)$:
\begin{enumerate}
\item
$a \in \outAB$ (hence $a \in \out_{Q_{\outAB}}$) and $s_B  \must{a}{}{B} s'_B$\\
then $((s_A,a q_A),(s_B,q_B)) \must{a}{}{\Omega(A) \stimes \Omega(B)} ((s_A,q_A),(s'_B,q_B))$.
\item
$a^\rhd \in \outAB^\rhd$ (hence $a^\rhd \in \inp_{Q_{\outAB}})$
and $s_A  \must{a}{}{A} s'_A$\\%M_A^\rhd)$\\
then $((s_A,q_A),(s_B,q_B)) \must{a^\rhd}{}{\Omega(A) \stimes \Omega(B)} ((s'_A,q_A a),(s_B,q_B))$. 
\end{enumerate}

Transitions which involve the output queue of $B$ are analogous.
\end{enumerate}

\noindent A detailed description of the form of the transitions of $\Omega(A) \stimes \Omega(B)$ is given in Appendix~\ref{sec:appendix-A}.

%%=======================================================

\section{Compatibility Notions}\label{sec:comp}

%In this work we focus on the interaction behavior of components and study their compatibility for
%synchronous and asynchronous communication. 
In this section we review our compatibility notions introduced in~\cite{tacas2010} for the synchronous
and in~\cite{DBLP:journals/corr/abs-1101-4731} for the asynchronous case.
%Our compatibility notions are inspired by the work of de Alfaro and Henzinger~\cite{Alfaro2001} based on the idea that whenever
%a component wants to issue an output $a$ then it should find its communication partner in a state
%where it is ready to accept $a$ as an input. 
%Our compatibility notions are inspired by de Alfaro and Henzinger~\cite{Alfaro2001} who have studied compatibility of
%interface automata in the case of synchronous communication.
For synchronous compatibility the idea is that whenever two synchronously cooperating components reach a state,
in which one of the components wants to send an output $a$, i.e., $a$ is enabled in the local state of the component,
and if this action $a$ belongs to the input actions of the other component, then the other component should be ready to receive $a$.
This means that the other component should be in a local state such that an outgoing transition labeled with $a$ exists.
An implicit assumption behind this definition is that outputs are autonomously selected by the sending component and therefore
its communication partner should accept (as an input) any possible output.

%For synchronous compatibility the idea is that whenever 
%a component wants to issue an output $a$ then its communication partner should be ready to accept $a$ as an input. 
%According to ~\cite{Alfaro2001,Alfaro2005} a communication error occurs if
%``in the product of two interface automata, one of the automata may produce an output action that is in the input alphabet of the other automaton,
%but is not accepted''. 
%Similarly, in the context of modal I/O-transition systems, the authors of~\cite{productlines} say:
%``An \emph{error state} is a state in which one component can output something that the other component might be unable to receive.''
%Similarly, Carmona and Kleijn~\cite{DBLP:journals/tcs/CarmonaK13} talk about message loss
%``when a component is not ready to receive as input a message sent to it.''
%multi-component environment,
%different synchronization strategies (blocking, non-blocking, ...) based on the synchronous product.
%This motivates our first compatibility notion:% for IOTSes, which is called \emph{strong synchronous compatibility}.

\begin{defi}[Strong synchronous compatibility]\label{def:syc}
  Two IOTSes $A$ and $B$ are \emph{strongly synchronously compatible},
  denoted by $A \syc B$, if they are composable and if
  for all reachable states $(s_A,s_B) \in \reach(A \stimes B)$,
  \begin{enumerate}
  \item $\forall a \in \out_A \cap \inp_B : \
      s_A \must{a}{}{A} s'_A \Longrightarrow
      \exists\ s_B  \must{a}{}{B} s'_B$,
 \item $\forall a \in \out_B \cap \inp_A : \
      s_B \must{a}{}{B} s'_B \Longrightarrow
      \exists\ s_A  \must{a}{}{A} s'_A$.
  \end{enumerate}
\end{defi} 

%This definition requires that IOTSes should work properly together in \emph{any}
%environment, in contrast to the ``optimistic'' approach of~\cite{Alfaro2001} %and~\cite{DBLP:conf/esop/LarsenNW07}, 
%in which the existence of a ``helpful'' environment to avoid error states is sufficient.
%For closed systems this makes no difference.
%%Note, that compatibility is not the same as deadlock-freeness.
In~\cite{tacas2010} we have introduced a weak version of compatibility
such that a component can delay the required input
and perform some internal actions before.
%We have shown in~\cite{DBLP:conf/tacas/BauerMSH10} that weak compatibility
We have shown in~\cite{tacas2010} that this fits well to weak refinement of component specifications in the sense
that weak refinement (in particular, weak bisimulation) preserves weak compatibility while it does not preserve strong compatibility. 
Refinement is, however, not a topic of this work.

\begin{defi}[Weak synchronous compatibility]\label{def:wsc}
  Two IOTSes $A$ and $B$ are \emph{weakly synchronously compatible},
  denoted by $A \wsc B$, if they are composable and if
  for all reachable states $(s_A,s_B) \in \reach(A \stimes B)$,
  \begin{enumerate}
  \item $\forall a \in \out_A \cap \inp_B : \
       s_A \must{a}{}{A} s'_A \Longrightarrow
      \exists\ s_B \  \xmust{\internal_B}{*}{B} \  \overline s_B \must{a}{}{B} s'_B$,
 \item $\forall a \in \out_B \cap \inp_A : \
      s_B \must{a}{}{B} s'_B \Longrightarrow
      \exists\ s_A \  \xmust{\internal_A}{*}{A} \  \overline s_A \must{a}{}{A} s'_A$,
  \end{enumerate}
\end{defi}

Now we turn to compatibility of asynchronously communicating components $A$ and $B$.
In this case outputs of a component are stored in a queue from which they can be consumed
by the receiver component. Therefore, in the asynchronous context, compatibility means that whenever a queue is not empty,
the receiver component must be ready to take (i.e.\ input) the next removable element from the queue.
Since we have enhanced components by output queues (rather than input queues) this idea can be easily formalized
by reduction to synchronous compatibility of the components $\Omega(A)$ and $\Omega(B)$.
Indeed, $\Omega(A)$ has an output $a$ enabled iff $a$ is the first element of the output queue of $A$
and the same holds symmetrically for $\Omega(B)$.

\begin{defi}[Strong and weak asynchronous compatibility]
Let $A$ and $B$ be two asynchronously composable I/O-transition systems.
$A$ and $B$ are \emph{strongly asynchronously compatible}, denoted by $A \ac B$,
if $\Omega(A) \syc \Omega(B)$.
$A$ and $B$ are \emph{weakly asynchronously compatible}, denoted by $A \wac B$,
if $\Omega(A) \wsc \Omega(B)$.
\end{defi}

\begin{exa}\label{ex:maker-user}
Fig.~\ref{fig:maker-user} shows the behavior of a \texttt{Maker} and a \texttt{User} process.
Here and in the subsequent drawings we use the following notations:
Initial states are denoted by $0$,
input actions $a$ are indicated by $a?$,
output actions $a$ by $a!$, and internal actions $a$ by $\tau_a$.
The maker expects some material from the
environment (input action \texttt{material}), constructs some item (internal action \texttt{make}), and then it signals either that the item is ready (output action \texttt{ready}) or that the production did fail (output action \texttt{fail}).
Both actions are shared with input actions of the user. When the user has
received the ready signal it uses the item (internal action \texttt{use}).
%The maker may also send \texttt{fail} to to the user to indicate that the construction of the item was not successful.  
\texttt{Maker} and \texttt{User} are weakly synchronously compatible but not strongly synchronously compatible.
The critical state in the synchronous product\linebreak
 \texttt{Maker} $\stimes$ \texttt{User} is $(2,1)$ 
which can be reached with the transitions

%$(0,0) \xmust{\mathrm{material}}{}{}\, (1,0) \xmust{\tau_{\mathrm{make}}}{}{}\, (2,0) \xmust{\tau_{\mathrm{ready}}}{}{}\, (0,1) \xmust{\mathrm{material}}{}{}\, (1,1) \xmust{\tau_{\mathrm{make}}}{}{}\, (2,1)$.

$(0,0) \xmust{\mathrm{material}}{}{}\, (1,0) \xmust{\mathrm{make}}{}{}\, (2,0) \xmust{\mathrm{ready}}{}{}\, (0,1) \xmust{\mathrm{material}}{}{}\, (1,1) \xmust{\mathrm{make}}{}{}\, (2,1)$.

\noindent In this state the maker wants to send \texttt{ready} or \texttt{fail} but the user must first perform its internal \texttt{use} action before it can receive the corresponding input.
The asynchronous composition \texttt{Maker} $\atimes$ \texttt{User} has infinitely many states
since the maker can be faster then the user.
We will see, as an application of the forthcoming results, %in particular Cor.~\ref{cor:io-sep},
that \texttt{Maker} and \texttt{User} are also weakly asynchronously compatible.
%Whenever a
%\texttt{ready} or \texttt{fail} message is on top of the output queue of \texttt{Maker}
%then the \texttt{User} can consume it, possibly after an internal \texttt{use} action.

\begin{figure}
  \centering
 \includegraphics[scale=0.7]{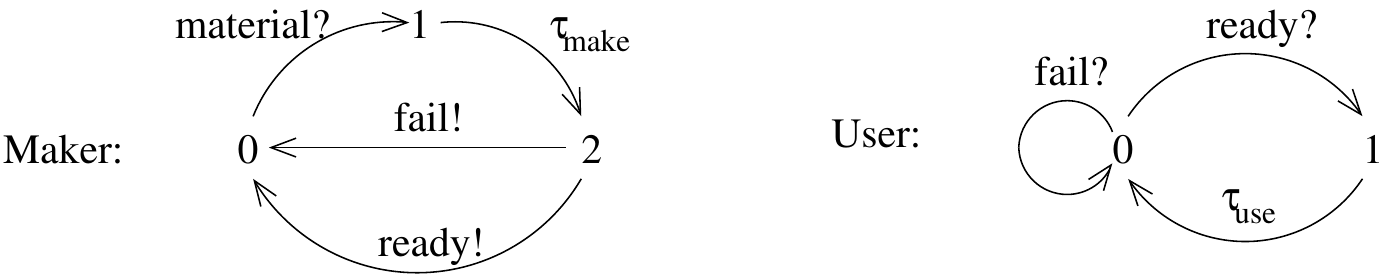}
  \caption{\texttt{Maker} and \texttt{User}}
  \label{fig:maker-user}
\end{figure}

\end{exa}

%It is, however, important to notice that the weak asynchronous compatibility notion adds much flexibility
%since it allows a receiver component to take available messages from the queue only after it
%has put its own issued messages in its output queue (since putting a message in a queue is an internal enqueue action of the form $a^\rhd$). 

%\vspace{4mm}
%TODO: Literatur Diskussion wo?
%
%
%For closed systems of finite IOTSes without internal actions the behavior described by their asynchronous composition coincides (up to naming conventions)
%with the operational model of communicating finite state machines (CFSMs) in~\cite{Brand-Zafiropulo}.
%In~\cite{Brand-Zafiropulo} a condition is formulated which requires that in a system of CFSMs executable receptions should be \emph{specified},
%which is just the strong version of the asynchronous compatibility notion used here.
%It is, however, important to notice that the weak asynchronous compatibility notion adds much flexibility
%since it allows a component to put first its own issued messages in its output queue (which are internal enqueue actions)
%before it takes available inputs already waiting in the other queue. An example for this situation will be provided later
%in Ex.~\ref{ex:wac}.

%=======================================================
%%=======================================================
\section{Relating Synchronous and Asynchronous Compatibility}\label{sec:relating}

As pointed out in Sect.~\ref{sec:intro}, it is generally undecidable whether two IOTSes are asynchronously compatible. 
In this section we study relationships between synchronous and asynchronous compatibility and conditions under which both are equivalent.
Under these conditions we can reduce asynchronous compatibility checking to synchronous compatibility checking which is decidable for finite state components.
%This is particularly motivated by the fact that for finite IOTSes reachability, and therefore synchronous (strong and weak)
%compatibility, are decidable which is in general not the case for asynchronous communication with unbounded FIFO-buffers.

\subsection{From Synchronous to Asynchronous Compatibility}\label{sec:sync2asynch}

\begin{figure}
  \centering
 \includegraphics[scale=0.7]{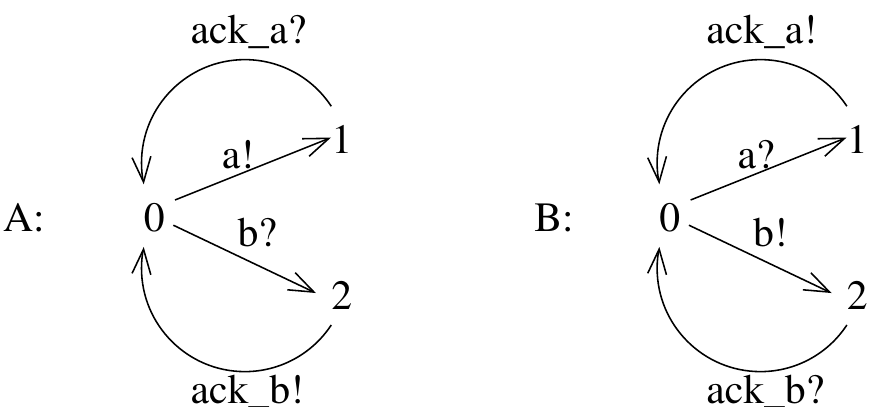}
  \caption{\texttt{A} $\syc$ \texttt{B} but \underline{not} \texttt{A} $\ac$ \texttt{B}}
  \label{fig:not-io-sep}
\end{figure}

We are interested in conditions  under which it is sufficient to check strong (weak) synchronous compatibility to ensure strong (weak) asynchronous compatibility.
In general this implication does not hold.
As an example consider the two IOTSes \texttt{A} and \texttt{B} in Fig.~\ref{fig:not-io-sep}.
Obviously, \texttt{A} and \texttt{B} are strongly synchronously compatible.
They are, however, neither strongly nor weakly asynchronously compatible since \texttt{A} may first put \texttt{a} in its output queue,
then \texttt{B} can output \texttt{b} in its queue and then both are blocked (\texttt{A} can only accept $\texttt{ack\_a}$
while \texttt{B} can only accept $\texttt{ack\_b}$).
%Obviously, $A$ and $B$ are also not weakly asynchronously compatible.
In Fig.~\ref{fig:not-io-sep} each IOTS has a state (the initial state) where a choice between an output and an input action is possible.
We will see (Cor.~\ref{cor:io-sep}) that if such situations are avoided synchronous compatibility implies asynchronous compatibility,
and we will even get more general criteria (Thm.~\ref{thm:sync2async}) for which the following property $\mathcal{P}$ is important.
%More generally, we will consider \emph{half-duplex} systems (see, e.g.,~\cite{DBLP:journals/iandc/CeceF05}),  in which always at least one
%output queue of the asynchronously communicating components is empty.

\noindent\textbf{Property} $\mathcal{P}$:
Let $A$ and $B$ be two asynchronously composable IOTSes. 
The asynchronous system $A \atimes B$ satisfies property $\mathcal{P}$ if
for each reachable state $((s_A,q_A),(s_B,q_B)) \in \reach(\Omega(A) \stimes \Omega(B))$ 
one of the following conditions holds:

\begin{enumerate}
\item[(i)]
$q_A = q_B = \emptyseq$ and $(s_A,s_B) \in \reach(A \stimes B)$. 
\item[(ii)]
$q_A = a_1 \ldots a_m \neq \emptyseq$ and $q_B = \emptyseq$ and there exists $r_A \in \states_A$ such that:\\
$(r_A,s_B) \in \reach(A \stimes B)$ and 
$r_A \Must{a_1}{}{A} \ldots \Must{a_m}{}{A} s_A$.
\item[(iii)]
$q_A = \emptyseq$ and $q_B = b_1 \ldots b_m \neq \emptyseq$ and there exists $r_B \in \states_B$ such that:\\
$(s_A,r_B) \in \reach(A \stimes B)$ and 
$r_B \Must{b_1}{}{B} \ldots \Must{b_m}{}{B} s_B$.
\end{enumerate}

To define the notation $\Must{a}{}{A}$, let $a \in \out_A \cap \inp_B$ and $F_A = \act_A \smallsetminus \shared(A,B)$ be the set of the free actions of $A$.
%To explain the notation $\Must{a}{}{A}$, let $X_A = \act_A \smallsetminus \shared(A,B)$ and $a \in \out_A \cap \inp_B$.
Then $s \Must{a}{}{A} s'$ holds if there exist transitions $s \xmust{F_A}{*}{A} \overline s \must{a}{}{A}\overline s' \xmust{F_A}{*}{A} s'$.
Recall from Sect.~\ref{sec:iolts} that $t \xmust{F_A}{*}{A} t'$ stands for a (possibly empty) sequence of transitions
involving only actions of $F_A$. Hence, $s \Must{a}{}{A} s'$
stands for a sequence of transitions
such that a single transition with $a \in \out_A \cap \inp_B$ is surrounded by arbitrary transitions in $A$ involving only free actions of  $A$. %$X_A$, i.e.\ internal actions of $A$
%or (free) output or input actions of $A$ not shared with $B$.
The notation $\Must{b}{}{B}$ is defined analogously.

Looking more closely to property $\mathcal{P}$ we see that it requires two things:
(a) In each reachable state of the asynchronous composition at least one of the two queues is empty, since one of the cases (i) - (iii) must always hold which entails emptiness of at least one queue.
(b) The state of the component where the output queue is not empty can be reached from a reachable state in the \emph{synchronous product}
by outputting the actions stored in the queue, possibly interleaved with free actions.
Part (a) specifies a well-known class of asynchronous systems called \emph{half-duplex} systems;
see, e.g.,~\cite{DBLP:journals/iandc/CeceF05}.

\begin{defi}\label{def:half-duplex}
Let $A$ and $B$ be two asynchronously composable IOTSes.
The asynchronous system $A \atimes B$ is \emph{half-duplex}, if for all reachable states
 $((s_A,q_A),(s_B,q_B)) \in \reach(\Omega(A) \stimes \Omega(B))$ it holds that $q_A = \emptyseq$ or  $q_B = \emptyseq$.
\end{defi}

It turns out that part (b) explained above is not really an extra condition. It is already a consequence of part (a), i.e., of being
a half-duplex system.
Hence, property $\mathcal{P}$ and being half-duplex are equivalent conditions, as shown in the crucial lemma below.
The direction (1) $\Rightarrow$ (2) is trivial since property $\mathcal{P}$ entails (a) from above, i.e., the half-duplex property.
The difficult direction is  (2) $\Rightarrow$ (1). The induction proof for this direction relies
on condition (3) of Lem.~\ref{lem:main} which is an easy consequence of being half-duplex.
A nice side-effect of Lem.~\ref{lem:main} is the equivalence (2) $\Leftrightarrow$ (3).
It shows how the half-duplex property can be decided, since (3) is decidable for finite state components.
This corresponds to a result in~\cite{DBLP:journals/iandc/CeceF05} where it is shown that membership to the class of half-duplex systems
is decidable.

\begin{lem}[Crucial lemma]\label{lem:main}
Let $A$ and $B$ be two asynchronously composable IOTSes. The following conditions are equivalent:
\begin{enumerate}
\item
 The asynchronous system $A \atimes B$ satisfies property $\mathcal{P}$.
\item
The asynchronous system $A \atimes B$ is half-duplex.
\item
For each reachable state $(s_A,s_B) \in \reach(A \stimes B)$
and each transitions $s_A \must{a}{}{A} s'_A$ and $s_B \must{b}{}{B} s'_B$
either $a \notin \out_A \cap \inp_B$ or $b \notin \out_B \cap \inp_A$.
% i.e. it is not possible that $A$ can issue an output for $B$ and $B$ can to issue an output for $A$ at the same time.
\end{enumerate}
\end{lem}

\proof
(1) $\Rightarrow$ (2) is trivial. (2) $\Rightarrow$ (3) is proved by contradiction: Assume (3) does not hold.
Then there exist a reachable state $(s_A,s_B) \in \reach(A \stimes B)$ and transitions
$s_A \must{a}{}{A} s'_A$ and $s_B \must{b}{}{B} s'_B$ such that
$a \in \out_A \cap \inp_B$ and $b \in \out_B \cap \inp_A$. Now we allow us a forward reference
to Lem.~\ref{lem:async2synch}, which shows
$((s_A,\emptyseq), (s_B,\emptyseq)) \in \reach(\Omega(A) \stimes \Omega(B))$.
Since $s_A \must{a}{}{A} s'_A$ we get a transition
$$((s_A,\emptyseq),(s_B,\emptyseq)) \must{a^\rhd}{}{\Omega(A) \stimes \Omega(B)} ((s'_A,a),(s_B,\emptyseq)).$$
Since $s_B \must{b}{}{B} s'_B$ we get a transition
$$((s'_A,a),(s_B,\emptyseq)) \must{b^\rhd}{}{\Omega(A) \stimes \Omega(B)} ((s'_A,a),(s'_B,b))$$
and therefore the system is not half-duplex.

The direction (3) $\Rightarrow$ (1) is proved by induction on the length of the derivation to reach
$((s_A,q_A),(s_B,q_B)) \in \reach(\Omega(A) \stimes \Omega(B))$. 
It  involves a complex case distinction on the form of the transitions in the asynchronous composition.
Interestingly only the case of transitions with enqueue actions needs the assumption (3).
The compete proof of (3) $\Rightarrow$ (1) is given in Appendix~\ref{sec:appendix-B}.
The interesting case in this proof is Case 5 (iii).
\qed

\begin{thm}[Synch2Asynch]\label{thm:sync2async}
Let $A$ and $B$ be two asynchronously composable IOTSes such that one (and hence all)
of the conditions in Lemma~\ref{lem:main} are satisfied. Then the following holds:
\begin{enumerate}
\item
 $A \syc B  \Longrightarrow A \ac B$.
\item
 $A \wsc B  \Longrightarrow A \wac B$. 
\end{enumerate}
\end{thm}

\proof
The proof uses Lem.~\ref{lem:main} for both cases.\\
(1) Assume $A \syc B$. We have to show  $\Omega(A) \syc \Omega(B)$. We prove condition (1) of Def.~\ref{def:syc}.
Condition (2) is proved analogously.\\
Let $((s_A,q_A),(s_B,q_B)) \in \reach(\Omega(A) \stimes \Omega(B))$,
$a \in \out_{\Omega(A)} \cap \inp_{\Omega(B)}$ and
$ (s_A,q_A) \must{a}{}{\Omega(A)} (s'_A,q'_A)$. Then $q_A$ has the form $a a_2 \ldots a_m$.
By assumption, $\Omega(A) \stimes \Omega(B)$ satisfies property $\mathcal{P}$. 
Hence, 
there exists $r_A \in \states_A$ such that $(r_A,s_B) \in \reach(A \stimes B)$
and  $r_A \Must{a}{}{A} \overline r_A \Must{a_2}{} \ldots \Must{a_m}{}{A} s_A$.
Thereby $r_A \Must{a}{}{A} \overline r_A$ is of the form
$r_A \xmust{F_A}{*}{A} s \must{a}{}{A} s' \xmust{F_A}{*}{A}  \overline r_A$.
Since $F_A$ involves only free actions of $A$ (not shared with $B$),
and since $(r_A,s_B) \in \reach(A \stimes B)$ we have that $(s,s_B) \in \reach(A \stimes B)$.
Now we can use the assumption $A \syc B$ which says that there exists $s_B  \must{a}{}{B} s'_B$.
Since $a \in \inp_B$, we get a transition $(s_B,q_B) \must{a}{}{\Omega(B)} (s'_B,q_B)$ and we are done.

\noindent(2) The weak case is a slight generalization of the proof of (1). The first part of the proof is the same
but then we use the assumption $A \wsc B$ which says that there exists 
$s_B \  \xmust{\internal_B}{*}{B} \  \overline s_B \must{a}{}{B} s'_B$ consisting of a sequence of 
internal transitions of $B$ followed by  $\overline s_B \must{a}{}{B} s'_B$ with $a \in \inp_B$.
Therefore we get transitions $(s_B,q_B) \  \xmust{\internal_B}{*}{\Omega(B)} \  (\overline s_B,q_B)$ 
$\must{a}{}{\Omega(B)} (s'_B,q_B)$
and, since  $\internal_B \subseteq \internal_{\Omega(B)}$, we are done. 
\qed

Note that the half-duplex property is not necessary for getting implication (1) and (2) of the last theorem.
An example would be two components $A$ and $B$ such that $\out_A \cap \inp_B = \{a\}$, $\out_B \cap \inp_A = \{b\}$
are the only actions, $A$ has one state and two looping transitions labeled with $a$ and $b$, and the same holds for $B$. 
Then $A$ and $B$ are synchronously and asynchronously strongly and weakly compatible, but the system is not half-duplex.
In fact, condition (3) of Lem.~\ref{lem:main} is violated.

We come back to our discussion at the beginning of this section where we have claimed that for I/O-transition systems
which do not show states where input and output actions are both enabled, % at the same time,
synchronous compatibility implies asynchronous compatibility. We must, however, be careful whether we consider the strong
or the weak case which leads us to two versions of I/O-separation.

\begin{defi}[I/O-separated transition systems]\label{def:io-sep}
Let $A$ be an IOTS.
\begin{enumerate}
\item
$A$ is called \emph{I/O-separated} if for all reachable states $s \in \reach(A)$ it holds:
If there exists a transition $s  \must{a}{}{A} s'$  with $a \in \out_A$ then there is no transition $s  \must{a'}{}{A} s''$  with $a' \in \inp_A$.

\item
$A$ is called \emph{observationally I/O-separated} if for all reachable states $s \in \reach(A)$ it holds:
If there exists a transition $s  \must{a}{}{A} s'$  with $a \in \out_A$ then there is no sequence of transitions
$s \  \xmust{\internal_A}{*}{A} \  \overline s_A \must{a'}{}{A} s''$ with $a' \in \inp_A$. 
\end{enumerate} 
\end{defi}

Obviously, observational I/O-separation implies I/O-separation but not the other way round;
cf. Ex.~\ref{ex:not-obs-io-sep}.
%Fig.~\ref{fig:not-obs-io-sep} shows an example of two IOTSes \texttt{A} and \texttt{B} which are 
%I/O-separated but not observationally I/O-separated.

\begin{lem}\label{lem:io-sep}
Let $A$ and $B$ be two asynchronously composable IOTSes.
\begin{enumerate}
\item
If $A$ and $B$ are I/O-separated and $A \syc B$, then
one (and hence all) of the conditions in Lemma~\ref{lem:main} are satisfied.
\item
If $A$ and $B$ are observationally I/O-separated and $A \wsc B$, then
one (and hence all) of the conditions in Lemma~\ref{lem:main} are satisfied. 
\end{enumerate}
\end{lem}

\proof
(1) By contradiction:
Assume condition (3) of Lem.~\ref{lem:main} does not hold. Then there are a reachable state $(s_A,s_B) \in \reach(A \stimes B)$
and transitions $s_A \must{a}{}{A} s'_A$ and $s_B \must{b}{}{B} s'_B$
such that $a \in \out_A \cap \inp_B$ and $b \in \out_B \cap \inp_A$.
Since $A \syc B$ there is a transition $s_B  \must{a}{}{B} s''_B$ with $a \in \inp_B$.
Therefore $B$ is not I/O-separated.

\noindent(2) is proved similarly by contradiction: Assume that condition (3) of Lem.~\ref{lem:main} does not hold.
This gives us again a reachable state $(s_A,s_B) \in \reach(A \stimes B)$
and transitions $s_A \must{a}{}{A} s'_A$ and $s_B \must{b}{}{B} s'_B$
such that $a \in \out_A \cap \inp_B$ and $b \in \out_B \cap \inp_A$.
Since $A \wsc B$ there exist transitions $s_B \  \xmust{\internal_B}{*}{B} \  \overline s_B \must{a}{}{B} s''_B$
with $a \in \inp_B$.
Therefore $B$ is not observationally I/O-separated.
%Hence, condition (3) of Lem.~\ref{lem:main} holds.
%Since  $A \wsc B$ we can apply Thm.~\ref{thm:sync2async}(2) and obtain $A \wac B$. \qed
\qed

The notion of I/O-separation appears in a more strict version, called \emph{input-separation},
in~\cite{DBLP:conf/monterey/HennickerJK08} and similarly as \emph{system without local mixed states}
in~\cite{DBLP:journals/iandc/CeceF05}. \cite{weiglhofer-diss} introduces internal choice labeled transition systems
which are particular versions of I/O-separated transition systems. The difference is that I/O-separation still allows
internal actions as an alternative to an input.
Part (1) of Lem.~\ref{lem:io-sep} can be considered as a generalization of Lemma 4 in~\cite{DBLP:conf/monterey/HennickerJK08} which has shown that input-separated IOTSes
which are strongly compatible and form a closed system are half-duplex.
This result was in turn a generalization of Thm. 35 in~\cite{DBLP:journals/iandc/CeceF05}.
Open systems and weak compatibility were not an issue in these approaches.
With Theorem~\ref{thm:sync2async} and Lemma~\ref{lem:io-sep} we get:
%Part (1) of Lem.~\ref{lem:io-sep} generalizes Lemma 4 in~\cite{DBLP:conf/monterey/HennickerJK08}
%to open systems and part (2) is a further generalization taking
%into account weak compatibilty.\footnote{Let us note that Lem. 4 in~\cite{DBLP:conf/monterey/HennickerJK08}
%was already a generalization of Thm. 35 in~\cite{DBLP:journals/iandc/CeceF05}.}

\begin{cor}\label{cor:io-sep}
Let $A$ and $B$ be two asynchronously composable IOTSes.
\begin{enumerate}
\item
If $A$ and $B$ are I/O-separated and $A \syc B$, then $A \ac B$.
\item
If $A$ and $B$ are observationally I/O-separated and $A \wsc B$, then $A \wac B$.
\end{enumerate}
\end{cor}

As an application of Cor.~\ref{cor:io-sep} we refer to Ex.~\ref{ex:maker-user}.
\texttt{Maker} and \texttt{User} are observationally I/O-separated, they
are weakly synchronously compatible and therefore, by Cor.~\ref{cor:io-sep}(2),
they are also weakly asynchronously compatible.

\begin{exa}\label{ex:not-obs-io-sep}
It may be interesting to note that part (2) of Cor.~\ref{cor:io-sep} and of Lem.~\ref{lem:io-sep} would not hold, if we would
only assume I/O-separation. Fig.~\ref{fig:not-obs-io-sep} shows two I/O-separated IOTSes \texttt{A} and \texttt{B}
with internal actions \texttt{i} and \texttt{j} resp., such that \texttt{A} and \texttt{B} are not  observationally I/O-separated.
 \texttt{A} and \texttt{B} are weakly synchronously compatible but not weakly asynchronously compatible
and the asynchronous system $A \atimes B$ is also not half-duplex.

\begin{figure}
  \centering
 \includegraphics[scale=0.7]{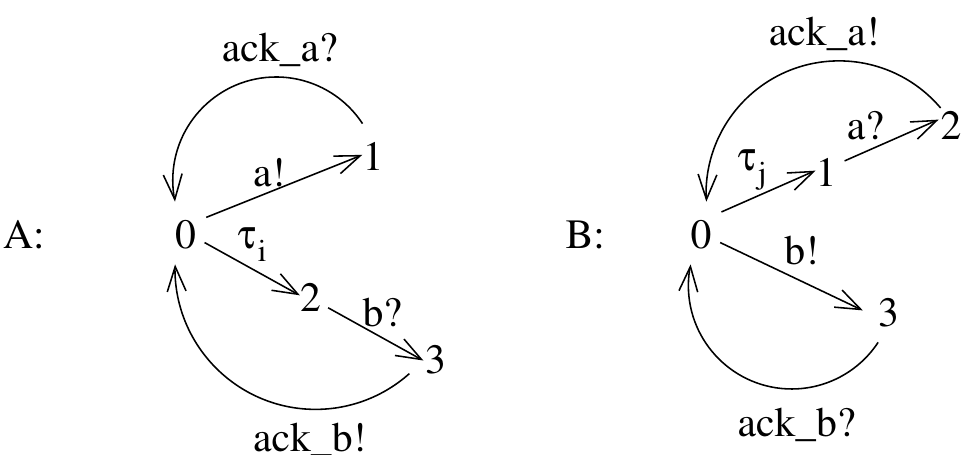}
 \caption{I/O-separated and $\texttt{A} \wsc \texttt{B}$ but \underline{not} $\texttt{A} \wac \texttt{B}$}
%  \caption{I/O-separated and $\wsc$ but not obs. I/O-separated and not $\wac$}
  \label{fig:not-obs-io-sep}
\end{figure}
\end{exa}

%=======================================================
%=======================================================

\subsection{From Asynchronous to Synchronous Compatibility}\label{sec:async2synch}

This section studies the other direction,
 i.e.\ whether asynchronous compatibility can imply
synchronous compatibility. It turns out that for the strong case this is indeed true without any further assumption
while for the weak case this holds under the equivalent conditions of Lem.~\ref{lem:main}.
In any case, we need for the proof the following lemma which shows that all reachable states
in the synchronous product are reachable in the asynchronous product with empty output queues.

\begin{lem}\label{lem:async2synch}
Let $A$ and $B$ be two asynchronously composable IOTSes. 
For any state $(s_A,s_B) \in \reach(A \stimes B)$, the state
$((s_A,\emptyseq), (s_B,\emptyseq))$ belongs to $\reach(\Omega(A) \stimes \Omega(B))$.
\end{lem}

\proof
The proof is straightforward by induction on the length of the derivation of $(s_A,s_B) \in \reach(A \stimes B)$.
It is given in the Appendix.
\qed

\begin{thm}[Asynch2Synch]\label{thm:async2sync}
For asynchronously composable IOTSes $A$ and $B$ it holds:
\begin{enumerate}
\item
 $A \ac B  \Longrightarrow A \syc B$.
\item
If one (and hence all) of the conditions in Lemma~\ref{lem:main} are satisfied, then
 $A \wac B  \Longrightarrow A \wsc B$. 
\end{enumerate}
\end{thm}

\proof
(1) Assume $A \ac B$, i.e. $\Omega(A) \syc \Omega(B)$. We have to show $A \syc B$.
We prove condition (1) of Def.~\ref{def:syc}. Condition (2) is proved analogously.\\
Let $(s_A,s_B) \in \reach(A \stimes B), a \in \out_A \cap \inp_B$ and $s_A \must{a}{}{A} s'_A$.
By Lem.~\ref{lem:async2synch}, $((s_A,\emptyseq), (s_B,\emptyseq))$ $\in \reach(\Omega(A) \stimes \Omega(B))$.
Since $s_A \must{a}{}{A} s'_A$, we have a transition in $\Omega(A) \stimes \Omega(B)$ with enqueue action for $a$:
$((s_A,\emptyseq), (s_B,\emptyseq)) \must{a^\rhd}{}{\Omega(A) \stimes \Omega(B)} ((s'_A,a), (s_B,\emptyseq))$ and it holds
$((s'_A,a), (s_B,\emptyseq)) \in \reach(\Omega(A) \stimes \Omega(B))$.
Then, there is a transition $(s'_A,a) \must{a}{}{\Omega(A)}$  $(s'_A,\emptyseq)$.
Since $\Omega(A) \syc \Omega(B)$ there must be a transition
$(s_B,\emptyseq) \must{a}{}{\Omega(B)} (s'_B,\emptyseq)$.
This transtion must be caused by a transition $ s_B \must{a}{}{B} s'_B$ and we are done.

\noindent(2) Assume $A \wac B$, i.e. $\Omega(A) \wsc \Omega(B)$. We have to show $A \wsc B$.
We prove condition (1) of Def.~\ref{def:wsc}. Condition (2) is proved analogously.\\
Let $(s_A,s_B) \in \reach(A \stimes B), a \in \out_A \cap \inp_B$ and $s_A \must{a}{}{A} s'_A$.
With the same reasoning as in case (1) we get
$((s'_A,a), (s_B,\emptyseq)) \in \reach(\Omega(A) \stimes \Omega(B))$ and we get  a transition
$(s'_A,a) \must{a}{}{\Omega(A)}$  $(s'_A,\emptyseq)$.
Since $\Omega(A) \wsc \Omega(B)$ there are transitions
$(s_B,\emptyseq)\  \xmust{\internal_{\Omega(B)}}{*}{\Omega(B)} \  $ $(\overline s_B,\overline q_B) \must{a}{}{\Omega(B)} (s'_B,\overline q_B)$.
Since internal transitions of $\Omega(B)$ do not involve any steps of $\Omega(A)$, we have
$((s'_A,a),  (\overline s_B,\overline q_B)) \in \reach(\Omega(A) \stimes \Omega(B))$.
Due to the assumption that the conditions in Lemma~\ref{lem:main} are satisfied,
$\Omega(A) \stimes \Omega(B)$ is half-duplex and therefore $\overline q_B$ must be
empty and the same holds for all intermediate queues reached by the transitions in
$(s_B,\emptyseq)\  \xmust{\internal_{\Omega(B)}}{*}{\Omega(B)} \  (\overline s_B,\overline q_B)$. 
Therefore no enqueue action can occur in these transitions.
Noticing that $\internal_{\Omega(B)} = \internal_B \cup (\out_B \cap \inp_A)^\rhd$,
we get $(s_B,\emptyseq)\  \xmust{\internal_B}{*}{\Omega(B)} \ $ $ (\overline s_B,\emptyseq) \must{a}{}{\Omega(B)} (s'_B,\emptyseq)$
and all these transtions must be induced by transitions
$s_B \  \xmust{\internal_B}{*}{B} \  \overline s_B \must{a}{}{B} s'_B$, i.e. we are done.
\qed

As a consequence of Thms.~\ref{thm:sync2async} and~\ref{thm:async2sync}
we see that under the equivalent conditions of Lem.~\ref{lem:main}, in particular when the asynchronous system is half-duplex,
(weak) synchronous compatibility is equivalent to (weak) asynchronous compatibility.

\begin{cor}[SynchIFFAsynch]\label{cor:syncIFFasync}
Let $A$ and $B$ be two asynchronously composable IOTSes such that one (and hence all)
of the conditions in Lemma~\ref{lem:main} are satisfied. Then the following holds:
\begin{enumerate}
\item
 $A \syc B  \Longleftrightarrow A \ac B$.
\item
 $A \wsc B  \Longleftrightarrow A \wac B$. 
\end{enumerate}
\end{cor}

%=======================================================
%=======================================================

\section{Weak Asynchronous Compatibility: The General Case}\label{sec:asynch-comp}

In this section we are interested in the verification of asynchronous compatibility
in the general case, where at the same time both queues of the communicating components may be not empty.
We focus here on \emph{weak} asynchronous compatibility since non-half duplex systems are
often weakly asynchronously compatible but not weakly synchronously compatible.\footnote{This is in contrast to the strong case
where strong asynchronous compatibility implies strong synchronous compatibility; see Thm.~\ref{thm:async2sync}(1).}
A simple example would be two components which both start to send a message to each other and after that
each component takes the message addressed to it from the buffer. Such a system would be weakly asynchronously compatible
but not weakly synchronously compatible.

\begin{exa}\label{ex:general-case}

Fig.~\ref{fig:weak-asynch} shows two IOTSes \texttt{MA} and \texttt{MB}
which produce items for each other. After reception of some material from the environment  (input action \texttt{materialA}),
\texttt{MA} produces
an item (internal action \texttt{makeA}) followed by either a signal that the item is ready for use
(output \texttt{readyA}) or a signal that the production did fail (output \texttt{failA}).
Whenever \texttt{MA} reaches its initial state  it can also accept an input \texttt{readyB} and then use the item produced by \texttt{MB}
(internal action \texttt{useB})
or it can accept a signal that the production of its partner did fail (input \texttt{failB}). 
The behavior of \texttt{MB} is analogous.
Note that \texttt{materialA} (resp.\ \texttt{materialB}) is a non-shared input action
of \texttt{MA} (resp.\ \texttt{MB}). They are open to the environment
after composition of \texttt{MA} and \texttt{MB}.
%Both IOTSes are observationally I/O-separated.
The asynchronous composition of \texttt{MA} and \texttt{MB} is not half-duplex;
both processes can produce and signal concurrently.
Clearly, the system is not weakly synchronously compatible. For instance,
the state (2,2) is reachable in the synchronous product and in this state
each of the two components wants to output an action but the other one is not able to synchronize with a corresponding input. We will prove
below that the system is weakly asynchronously compatible. 
Let us note that the system considered here is neither synchronizable in the sense of~\cite{DBLP:conf/facs2/OuederniSB13} nor desynchronizable in the sense of~\cite{DBLP:journals/scp/BeoharC14,beohar-diss}.
The reason is simple: The synchronous system is blocked in state (2,2) while the asynchronous
system can always proceed with putting messages in the buffers, consuming them and firing transitions for the free actions open to the environment. Therefore there cannot be a branching bisimulation between
the synchronous and the asynchronous versions of the system as required for synchronizability
in~\cite{DBLP:conf/facs2/OuederniSB13} and for desynchronizability in~\cite{DBLP:journals/scp/BeoharC14,beohar-diss}.

\begin{figure}
  \centering
 \includegraphics[scale=0.7]{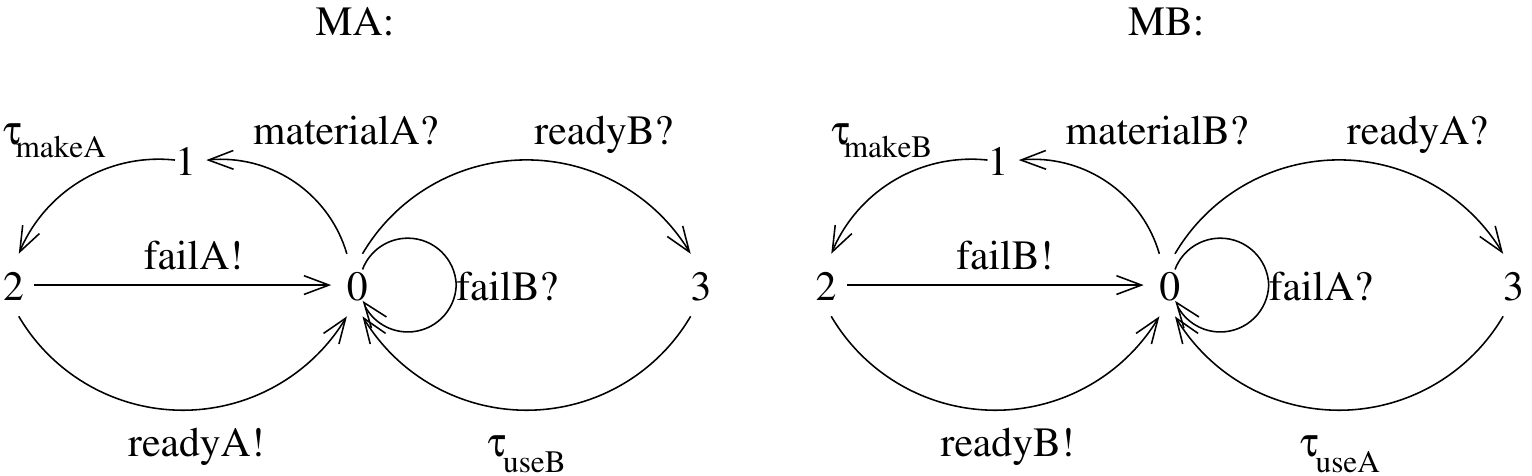}
  \caption{$\texttt{MA} \wac \texttt{MB}$ but \underline{not} $\texttt{MA} \wsc \texttt{MB}$.}
  \label{fig:weak-asynch}
\end{figure}

\end{exa}

In general, the problem of weak asynchronous compatibility is undecidable due to potentially unbounded message queues.
For instance, in the asynchronous composition of \texttt{MA} and \texttt{MB} in Ex.~\ref{ex:general-case} both output queues can grow without upper bound.
In Sect.~\ref{sec:criterion} we develop a criterion for proving weak asynchronous compatibility  in the general case
(allowing for non half-duplex systems with unbounded message queues).
The criterion is decidable if the underlying IOTSes are finite. In Sect.~\ref{sec:completeness} we investigate  properties under which the criterion is even complete, i.e.\ if the criterion is not satisfied, then the system is not weakly asynchronously compatible.

\subsection{A Criterion for Weak Asynchronous Compatibility}\label{sec:criterion}

Let $A$ and $B$ be two asynchronously composable IOTSes.
The idea for proving weak asynchronous compatibility of $A$ and $B$ is again to use synchronous products, but not the standard synchronous composition of $A$ and $B$ but variants of it.
First we focus only on one direction of compatibility concerning the outputs of $A$ which should
be received by $B$. Due to the weak compatibility notion $B$ can, before it takes an input message, execute internal actions.
In particular, it can put outputs directed to $A$ in its output queue. (Remember that enqueue actions are internal). To simulate these autonomous enqueue actions in a \emph{synchronous} product with $A$,
 we consider
the renamed version $B_{\outBA}^\rhd$ of $B$ where all actions $b \in \outBA = \out_B \cap \inp_A$ are renamed to $b^\rhd$.
Thus they become non-shared actions which can be freely executed in the synchronous product of $A$ and $B_{\outBA}^\rhd$
(just as the enqueue actions $b^\rhd$ in the asynchronous product of $A$ and $B$). 
At the same time all previously shared input actions of $A$ become free.
Now we require that in each reachable state of  the \emph{synchronous} product $A \stimes B_{\outBA}^\rhd$
if $A$ wants to send an output $a$ addressed to $B$ then $B_{\outBA}^\rhd$ can execute some internal actions and/or free output actions  $b^\rhd \in \outBA^\rhd$ before it accepts $a$.
This idea is formalized in the following condition (a).
A symmetric condition concerning the compatibility in the direction from $B$ to $A$  is formalized in condition (b).        

\begin{enumerate}
\item[(a)]
For all reachable states $(s_A,s_B) \in \reach(A \stimes \renBA)$,
$\forall a \in \outAB = \out_A \cap \inp_B : \
       s_A \must{a}{}{A} s'_A \Longrightarrow
      \exists\ s_B \  \xmust{\internal_{B}\cup\outBA^\rhd}{*}{\renBA} \  \overline s_B \must{a}{}{B} s'_B$.\footnote{Note that $\internal_{B} = \internal_{\renBA}$ and $\overline s_B \must{a}{}{B} s'_B$ is equivalent to $\overline s_B \must{a}{}{\renBA} s'_B$, since  $a \in \out_A \cap \inp_B$ is not renamed.} 

\item[(b)]
For all reachable states $(s_A,s_B) \in \reach(\renAB \stimes B)$,
$\forall b \in \outBA = \out_B \cap \inp_A : \
       s_B \must{b}{}{B} s'_B \Longrightarrow
      \exists\ s_A \  \xmust{\internal_{A}\cup\outAB^\rhd}{*}{\renAB} \  \overline s_A \must{b}{}{A} s'_A$.
\end{enumerate}

\begin{nota} We write $A \dashrightarrow \renBA$ if condition (a) holds and
$B \dashrightarrow \renAB$ if condition (b) holds.
\end{nota}

We call the conditions $A \dashrightarrow \renBA$ and $B \dashrightarrow \renAB$
the \emph{WAC-criterion} since they are sufficient for weak asynchronous compatibility.
%The next theorem shows that for any two asynchronously composable IOTSes $A$ and $B$,
%if $A \dashrightarrow \renBA$ and $B \dashrightarrow \renAB$ holds 
%then $A$ and $B$ are weakly asynchronously compatible.

\begin{thm}[WAC-criterion]\label{thm:general-case}
Let $A$ and $B$ be two asynchronously composable IOTSes such that
$A \dashrightarrow \renBA$ and $B \dashrightarrow \renAB$ holds.
Then $A$ and $B$ are weakly asynchronously compatible, i.e.\ $A \wac B$.
\end{thm}

The proof of this theorem needs an auxiliary, technical lemma
which establishes a relationship between the reachable states of the asynchronous composition of $A$ and $B$
and the reachable states considered in
the synchronous products $A \stimes \renBA$ and $\renAB \stimes B$ respectively.  
%  In contrast to property $\mathcal{P}$ they are generally valid (without half-duplex assumption).

\begin{lem}\label{lem:general-case}
For any two asynchronously composable IOTSes $A$ and $B$ it holds that $A$ and $\renBA$ as well as $\renAB$ and $B$ are synchronously composable and \emph{both} of the following two properties $\mathcal{Q}_A$ and $\mathcal{Q}_B$ are satisfied.

\noindent\textit{Property} $\mathcal{Q}_A$:
For each reachable state $((s_A,q_A),(s_B,q_B)) \in \reach(\Omega(A) \stimes \Omega(B))$
one of the following two conditions holds:

\begin{enumerate}
\item[(i)]
$q_A = \emptyseq$ and $(s_A,s_B) \in \reachBA$,
\item[(ii)]
$q_A = a_1 \ldots a_m \neq \emptyseq$ and there exists $r_A \in \states_A$ such that:\\
$(r_A,s_B) \in \reachBA$ and 
$r_A \xtRarrA{a_1} \ldots \xtRarrA{a_m} s_A$.\\
The notation $s \xtRarrA{a} s'$ stands for an arbitrary sequence of transitions in $A$ 
which contains exactly one transition with an output action in $\out_A \cap \inp_B$ and
this output action is $a$. 
\end{enumerate}

\noindent\textit{Property} $\mathcal{Q}_B$:
For each reachable state $((s_A,q_A),(s_B,q_B)) \in \reach(\Omega(A) \stimes \Omega(B))$
one of the following two conditions holds:

\begin{enumerate}
\item[(i)]
$q_B = \emptyseq$ and $(s_A,s_B) \in \reachAB$, 
\item[(ii)]
$q_B = b_1 \ldots b_m \neq \emptyseq$ and there exists $r_B \in \states_B$ such that:\\
$(s_A,r_B) \in \reachAB$ and 
$r_B \xtRarrB{b_1} \ldots \xtRarrB{b_m} s_B$.\\
The notation $\xtRarrB{b}$ is defined analogously to $\xtRarrA{a}$.
\end{enumerate}
\end{lem}

%\reachBA  \reachBA
%\reachAB  \reachAB

\proof
Since $A$ and $B$ are  asynchronously composable they are synchronously composable and
$\shared(A,B)^\rhd \cap (\act_A \cup \act_B) = \emptyset$.
Hence, $A$ and $\renBA$ as well as $\renAB$ and $B$ are synchronously composable.
  
The initial state $((\start_A,\emptyseq),(\start_B,\emptyseq))$ satisfies $\mathcal{Q}_A$ and $\mathcal{Q}_B$.
Then we consider transitions
$$ ((s_A,q_A),(s_B,q_B)) \must{a}{}{\Omega(A) \stimes \Omega(B)} ((s'_A,q'_A),(s'_B,q'_B))$$
%$$(*)\hspace{5mm} ((s_A,q_A),(s_B,q_B)) \must{a}{}{\Omega(A) \stimes \Omega(B)} ((s'_A,q'_A),(s'_B,q'_B))$$
and
show that if $ ((s_A,q_A),(s_B,q_B))$ satisfies  $\mathcal{Q}_A$ ($\mathcal{Q}_B$ resp.) then
$((s'_A,q'_A),(s'_B,q'_B))$ satisfies  $\mathcal{Q}_A$ ($\mathcal{Q}_B$ resp.). 
Then the result follows by induction on the length of the derivation to reach
$((s_A,q_A),(s_B,q_B)) \in \reach(\Omega(A) \stimes \Omega(B))$.
The complete proof is given in the Appendix.
%The proof is performed by case distinction
%on the form of the action $a$ and is given in the Appendix.
%taking into account the case distinction after Def.~\ref{def:asynch-composition}.
%Then the result follows by induction on the length of the sequence of transitions to reach an arbitrary state
%$((s_A,q_A),(s_B,q_B)) \in \reach(\Omega(A) \stimes \Omega(B))$. 
%The complete proof is given in the Appendix.
\qed

Property $\mathcal{Q}_A$(i) expresses that whenever a global state $((s_A,\emptyseq),(s_B,q_B))$
is reachable in the asynchronous composition of $A$ and $B$,
then $(s_A,s_B)$ is already reachable in the synchronous composition $A \stimes \renBA$.
Property $\mathcal{Q}_A$(ii) expresses that whenever a global state $((s_A,q_A),(s_B,q_B))$
with $q_A \neq \emptyseq$ is reachable in the asynchronous composition of $A$ and $B$
there exists a state $r_A$ of $A$ such that $(r_A,s_B)$ is reachable in the synchronous composition
$A \stimes \renBA$ and the local control state $s_A$ of $A$ 
can be reached from $r_A$
by outputting the actions stored in the queue, possibly interleaved with arbitrary other actions of $A$ which are not output actions directed to $B$. Properties $\mathcal{Q}_B$(i) and (ii) are the symmetric properties concerning the output queue of $B$.

The properties $\mathcal{Q}_A$ and $\mathcal{Q}_B$ have a pattern similar to property $\mathcal{P}$ in Sect.~\ref{sec:sync2asynch} which has related reachable states of the asynchronous composition of $A$ and $B$
with reachable states in the synchronous product $A \stimes B$. Such a relation was only possible under the half-duplex assumption while $\mathcal{Q}_A$ and $\mathcal{Q}_B$ are generally valid. The intuitive reason is that $A \stimes \renBA$ as well
as $\renAB \stimes B$ can have significantly more reachable states than  $A \stimes B$.
This is demonstrated in Ex.~\ref{ex:not-complete} and this is also the reason why
our proof technique is in general not complete; see Sect.~\ref{sec:completeness}.
We are now prepared to prove  Thm.~\ref{thm:general-case}.
%This is caused by the fact that
%%In fact, the essential difference between $A \stimes \renBA$ and $A \stimes B$ is that
%actions belonging to $\outBA = \out_B \cap \inp_A$ must synchronize in $A \stimes B$ since they are shared
%actions of $A$ and $B$. But after renaming $b$ to $b^\rhd$, the actions $b^\rhd$ with $b \in \outBA$
%become non-shared output actions of $\renBA$ which can be freely executed in the synchronous product
%$A \stimes \renBA$.
%At the same time the actions $b \in \outBA = \out_B \cap \inp_A$
%become non-shared input actions of $A$ which can also be freely executed in the synchronous product of $A$ and $B_{\outBA}^\rhd$.
%The fact that $A \stimes \renBA$ and $\renAB \stimes B$ may have more reachable states than  $A \stimes B$
%is also the reason that our proof technique is in general not complete; see the discussion in Sect.~\ref{sec:completeness}.
%We are now prepared to prove Th.~\ref{thm:general-case}.

%\begin{thm}\label{thm:general-case}
%Let $A$ and $B$ be two asynchronously composable IOTSes such that\linebreak
%$A \dashrightarrow \renBA$ and $B \dashrightarrow \renAB$ holds.
%Then $A$ and $B$ are weakly asynchronously compatible, i.e.\ $A \wac B$.
%\end{thm}

\textit{Proof of Theorem~\ref{thm:general-case}:}
%The proof relies on Lem.~\ref{lem:general-case}.
By definition of $A \wac B$ we have to show $\Omega(A) \wsc \Omega(B)$.
We prove condition (1) of Def.~\ref{def:wsc}. Condition (2) is proved analogously.\\
Let $((s_A,q_A),(s_B,q_B)) \in \reach(\Omega(A) \stimes \Omega(B))$.
To prove condition (1) of Def.~\ref{def:wsc} we assume given $a \in \out_{\Omega(A)} \cap \inp_{\Omega(B)}$ and
$ (s_A,q_A) \must{a}{}{\Omega(A)}$ $(s'_A,q'_A)$ and we must show that there exist transitions
$(s_B,q_B) \  \xmust{\internal_{\Omega(B)}}{*}{\Omega(B)} \  (\overline s_B,\overline q_B) \must{a}{}{\Omega(B)} (s'_B,q'_B)$.

By the assumption, $q_A$ must have the form $a a_2 \ldots a_m$.
By Lem.~\ref{lem:general-case}, property $\mathcal{Q}_A$(ii) holds for $((s_A,q_A),(s_B,q_B))$.
Hence, there exists $r_A \in \states_A$ such that
$(r_A,s_B) \in \reachBA$ and 
$r_A \xtRarrA{a} \overline r_A \xtRarrA{a_2} \ldots \xtRarrA{a_m} s_A$.
Thereby $r_A \xtRarrA{a} \overline r_A$ is of the form
$r_A \xmust{Y_A}{*}{A} s \must{a}{}{A} s'$  $\xmust{Y_A}{*}{A}  \overline r_A$
with $a \in \out_{\Omega(A)} \cap \inp_{\Omega(B)} = \out_A \cap \inp_B =  \outAB$ and $Y_A$ involves no action in $\outAB$.
Since $\outAB$ are the only shared actions of $A$ and $\renBA$,
 the transitions in $r_A \xmust{Y_A}{*}{A} s$ induce transitions in
$A \stimes\renBA$ without involving $\renBA$. 
Therefore, since $(r_A,s_B) \in \reachBA$, we get
$(s,s_B) \in \reachBA$.
Now we can use the assumption $A \dashrightarrow \renBA$ which says that there exists
a sequence of transitions
$$s_B \  \xmust{\internal_{B}\cup\outBA^\rhd}{*}{\renBA} \  \overline s_B \must{a}{}{B} s'_B.$$
The actions in $\internal_{B}\cup\outBA^\rhd$ are internal actions of $\Omega(B)$ such that we
get transitions
$$(s_B,q_B) \  \xmust{\internal_{\Omega(B)}}{*}{\Omega(B)} \  (\overline s_B,\overline q_B) \must{a}{}{\Omega(B)} (s'_B,\overline q_B)$$ 
where $\overline q_B$ extends $q_B$
according to the elements that have been enqueued with actions in $\outBA^\rhd$.
Thus $\Omega(B)$ accepts $a$, possibly after some internal actions, and we are done. 
\qed

\begin{exa}\label{exa:general}
To apply Thm.~\ref{thm:general-case} to Ex.~\ref{ex:general-case} we have to prove
$\texttt{MA} \dashrightarrow \texttt{MB}^\rhd_{\{\texttt{readyB,failB}\}}$
and
$\texttt{MB} \dashrightarrow \texttt{MA}^\rhd_{\{\texttt{readyA,failA}\}}$.
For the former case, Fig.~\ref{fig:compatibility-checkMA} shows the IOTS \texttt{MA} and the
IOTS $\texttt{MB}^\rhd_{\{\texttt{readyB,failB}\}}$ obtained by renaming of its outputs.
We will check only this case, the other one is analogous.
We have to consider the reachable states in the synchronous product
$\texttt{MA} \stimes \texttt{MB}^\rhd_{\{\texttt{readyB,failB}\}}$
and when an output \texttt{readyA} or \texttt{failA} is possible in \texttt{MA}.
These states are (2,0), (2,1) and (2,2) since \texttt{materialA}, \texttt{materialB} are non-shared input actions
and \texttt{makeA}, \texttt{makeB} are internal actions.
(Note that state (2,3) is not reachable in $\texttt{MA} \stimes \texttt{MB}^\rhd_{\{\texttt{readyB,failB}\}}$
because \texttt{readyA} is a shared action
of \texttt{MA} and $\texttt{MB}^\rhd_{\{\texttt{readyB,failB}\}}$.)

In state (2,0) any output \texttt{readyA} or \texttt{failA}
is immediately accepted by
$\texttt{MB}^\rhd_{\{\texttt{readyB,failB}\}}$.
In state (2,1), $\texttt{MB}^\rhd_{\{\texttt{readyB,failB}\}}$
can perform first the internal action \texttt{makeB}, then the free output action
$\texttt{readyB}^\rhd$ or $\texttt{failB}^\rhd$ and then it can accept the input \texttt{readyA} or \texttt{failA}.
In state (2,2), $\texttt{MB}^\rhd_{\{\texttt{readyB,failB}\}}$
can perform the free output action
$\texttt{readyB}^\rhd$ or $\texttt{failB}^\rhd$ and then accept the input.  
Therefore we have shown $\texttt{MA} \wac \texttt{MB}$. 
Note that with the free output actions we have simulated in the synchronous product the (internal) enqueue actions $\texttt{readyB}^\rhd$
and $\texttt{failB}^\rhd$ 
that can be executed by \texttt{MB} in the asynchronous composition.\footnote{Our technique would also work for the non-synchronizable system example  in~\cite{DBLP:conf/facs2/OuederniSB13}, Fig. 4.}

\begin{figure}
  \centering
 \includegraphics[scale=0.7]{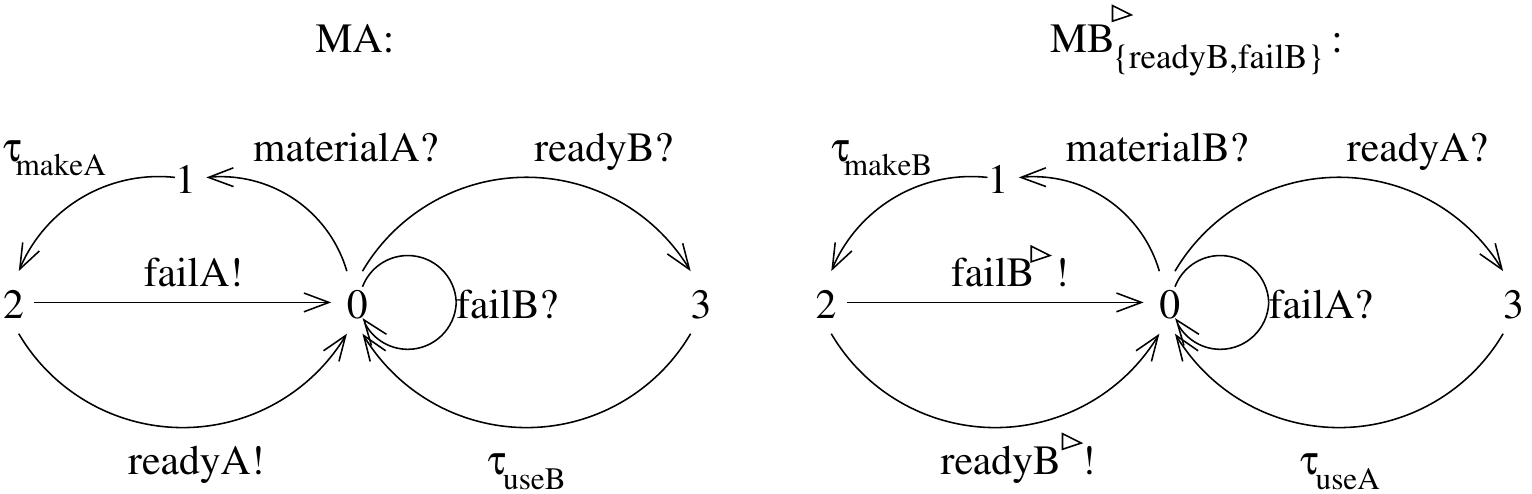}
  \caption{Compatibility check: $\texttt{MA} \dashrightarrow \texttt{MB}^\rhd_{\{\texttt{readyB,failB}\}} $}
  \label{fig:compatibility-checkMA}
\end{figure}
\end{exa}

%==========================================

\subsection{On the Completeness of the Compatibility Criterion}\label{sec:completeness}

%The compatibility criterion of the last section did rely on the two conditions (a) and (b)
%required for all reachable states  $(s_A,s_B) \in \reach(A \stimes \renBA)$ and 
%$(s_A,s_B) \in \reach(\renAB \stimes B)$ resp.
%Let us have a closer look to the reachable states $(s_A,s_B) \in \reach(A \stimes \renBA)$.

The compatibility criterion of the last section relies on the two conditions (a) and (b)
required for all reachable states  of $A \stimes \renBA$ and  $\renAB \stimes B$ respectively.
If $A$ and $B$ are finite then the compatibility criterion is decidable while weak
asynchronous compatibility is, in general, not decidable.
Hence the compatibility criterion cannot be complete.
In this section we first discuss in which situations it can happen that the compatibility criterion
is not necessary for weak asynchronous compatibility and then we establish a condition
under which the compatibility criterion is even complete for proving or disproving weak asynchronous
compatibility.

\begin{exa}\label{ex:not-complete}
The following very simple example illustrates the issue. %discussion.
We consider two components $A$ and $B$ such that
$\inp_A = \{b\}$, $\out_A = \{a\}$, $\internal_A = \emptyset$ and
$\inp_B = \{a\}$, $\out_B = \{b\}$, $\internal_B = \emptyset$.
The transitions of $A$ are shown in Fig.~\ref{fig:compatibility-check}.
The component $B$ and hence $\renBA$ has no transitions; i.e.\ their actions are never enabled.
Then it is trivial that $A$ and $B$ are weakly asynchronously compatible, since in the asynchronous composition
$A$ will never receive a message from $B$, i.e.
$\Omega(A) \stimes \Omega(B)$ will never reach a state
$(s_A,q_A),(s_B,q_B)$ with $s_A = 1$.
Therefore $A$ will never put $a$ in its output buffer.
However, our condition (a), $A \dashrightarrow \renBA$, is not satisfied since
$b$ is a free input action of $A$ in $A \stimes \renBA$ and therefore the state
 $(1,0)$ is reachable in $A \stimes \renBA$.
Then $A \dashrightarrow \renBA$ would require that
$\renBA$ is able to receive $b$ in its state 0 which is not the case.
%However, $(1,0)$ is reachable in $A \stimes \renBA$ (since $b$ is not a shared action of  $A$ and $\renBA$),
%there is a transition $1 \must{a}{}{A} 0$ with output $a$, but $\renBA$ can not input $a$.  
%Hence condition (a), i.e.\ $A   \dashrightarrow \renBA$, is not satisfied
%though $A$ and $B$ are weakly asynchronously compatible.

\begin{figure}
  \centering
 \includegraphics[scale=0.7]{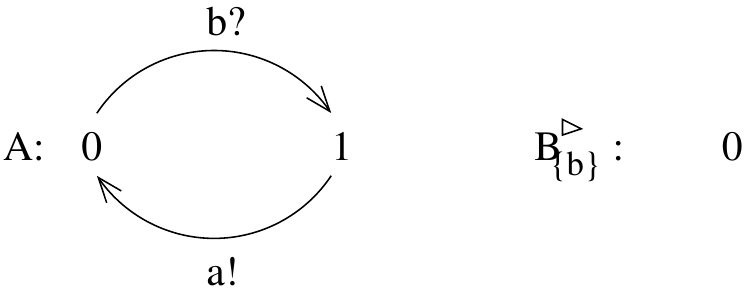}
  \caption{Compatibility check $\texttt{A} \dashrightarrow \texttt{B}^\rhd_{\{\texttt{b}\}}$ fails but $\texttt{A} \wac \texttt{B}$.}
  \label{fig:compatibility-check}
\end{figure}
\end{exa}

The problem encountered in Ex.~\ref{ex:not-complete} is that $A \stimes \renBA$ (or, symmetrically, $\renAB \stimes B$) may have more reachable states
than necessary to be considered in the asynchronous composition $A \atimes B$.
These states are reached by open inputs in  $A \stimes \renBA$ (or  $\renAB \stimes B$) which are never served in the asynchronous composition
where the inputs are shared actions.
More precisely, our conjecture is that the criterion of Thm.~\ref{thm:general-case}
may not be complete only if either

\begin{enumerate}
\item[(i)]
there are states $(s_A,s_B)$ reachable in 
$A \stimes \renBA$  such that $A$ has an output in state $s_A$ but the local state $s_A$ is not reachable in the asynchronous composition with $B$
and hence irrelevant for proving asynchronous compatibility in the direction from $A$ to $B$,
or 
\item[(ii)]
there are states $(s_A,s_B)$ reachable in $\renAB \stimes B$
 such that $B$ has an output in state $s_B$ but the local state $s_B$ is not reachable in the asynchronous composition with $A$
and hence irrelevant for proving asynchronous compatibility in the direction from $B$ to $A$.
\end{enumerate}

The subsequent theorem shows that our conjecture is right.
It relies on the definition of locally reachable states.

\begin{defi}
Let $A$ and $B$ be two synchronously composable IOTSes.
A state $s_A$ of $A$ is \emph{locally reachable} in $A \otimes B$,
if there exists a state $s_B$ of $B$ such that $(s_A,s_B) \in \reach(A \stimes B)$.
Local reachability for states of $B$ is defined analogously.
%Similarly, a state $s_B$ of $B$ is \emph{locally reachable} in $A \otimes B$,
%if there exists a state $s_A$ of $A$ such that $(s_A,s_B) \in \reach(A \stimes B)$. 
\end{defi}

\begin{thm}[Completeness criterion]\label{thm.completenss}
Let $A$ and $B$ be two asynchronously composable IOTSes such that
the following two properties  $\mathcal{X}_A$ and $\mathcal{X}_B$ are satisfied.

%\noindent\textit{Property} $\mathcal{X}_A$:
%For all reachable states  $(s_A,s_B) \in \reach(A \stimes \renBA)$
%for which a transition $s_A \must{a}{}{A} s'_A$ exists with $a \in \outAB$,
%it holds that there exist $s_B', q_B$ such that
%$((s_A,\emptyseq),(s_B',q_B)) \in \reach(\Omega(A) \stimes \Omega(B))$.
%
%\noindent\textit{Property} $\mathcal{X}_B$:
%For all reachable states  $(s_A,s_B) \in \reach(\renAB \stimes B)$
%for which a transition $s_B \must{b}{}{B} s'_B$ exists with $b \in \outBA$,
%it holds that there exist $s_A', q_A$ such that
%$((s_A',q_A),(s_B,\emptyseq)) \in \reach(\Omega(A) \stimes \Omega(B))$.

\noindent\textit{Property} $\mathcal{X}_A$:
For any state  $s_A$ of $A$ for which a transition $s_A \must{a}{}{A} s'_A$ exists with $a \in \outAB$
the following holds:
If $s_A$ is locally reachable in $A \stimes \renBA$ then $(s_A,\emptyseq)$
is locally reachable in $\Omega(A) \stimes \Omega(B)$.\footnote{In other words, 
there exist $s_B, q_B$ such that
$((s_A,\emptyseq),(s_B,q_B)) \in \reach(\Omega(A) \stimes \Omega(B))$.}

\noindent\textit{Property} $\mathcal{X}_B$:
For any state  $s_B$ of $B$ for which a transition $s_B \must{b}{}{B} s'_B$ exists with $b \in \outBA$
the following holds:
If $s_B$ is locally reachable in $\renAB \stimes B$ then $(s_B,\emptyseq)$
is locally reachable in $\Omega(A) \stimes \Omega(B)$.

Then 
$A \dashrightarrow \renBA$ and $B \dashrightarrow \renAB$ holds
if, and only if, $A$ and $B$ are weakly asynchronously compatible, i.e.\ $A \wac B$.
\end{thm}

\proof
Taking into account Thm.~\ref{thm:general-case}, it remains to show that
under the assumptions $\mathcal{X}_A$ and $\mathcal{X}_B$ we have that $A \wac B$
implies  $A \dashrightarrow \renBA$ and $B \dashrightarrow \renAB$.
Let $A \wac B$, i.e. $\Omega(A) \wsc \Omega(B)$. We show that then $A \dashrightarrow \renBA$ holds.
The proof of $B \dashrightarrow \renAB$ is analogous.

Let $(s_A,s_B) \in \reach(A \stimes \renBA)$.
To show $A \dashrightarrow \renBA$ we assume $a \in \outAB = \out_A \cap \inp_B$ and
$s_A \must{a}{}{A} s'_A$.
According to the meaning of $A \dashrightarrow \renBA$ (as defined in the notation before), we have to show that there exist transitions
$$ (*) s_B \  \xmust{\internal_{B}\cup\outBA^\rhd}{*}{\renBA} \  \overline s_B \must{a}{}{B} s'_B.$$
Since  $\mathcal{X}_A$ is valid, there exist $s_B, q_B$ such that
$((s_A,\emptyseq),(s_B,q_B)) \in \reach(\Omega(A) \stimes \Omega(B))$.
The transition $s_A \must{a}{}{A} s'_A$ induces an enqueue transition
$((s_A,\emptyseq),(s_B,q_B))$ $\must{a^\rhd}{}{\Omega(A) \stimes \Omega(B)}$ $((s'_A,a),(s_B,q_B))$
such that $a$ is the only element in the output queue of $A$.
Obviously, $((s'_A,a),(s_B,q_B)) \in \reach(\Omega(A) \stimes \Omega(B))$ and  there is a transition $(s'_A,a) \must{a}{}{\Omega(A)}$  $(s'_A,\emptyseq)$.
Since $\Omega(A) \wsc \Omega(B)$ there are transitions
$$(s_B,q_B) \  \xmust{\internal_{\Omega(B)}}{*}{\Omega(B)} \  (\overline s_B,\overline q_B) \must{a}{}{\Omega(B)} (s'_B,\overline q_B).$$
Since $\internal_{\Omega(B)} = \internal_{B}\cup\outBA^\rhd$ and $a \in \inp_{\Omega(B)} = \inp_B$
there are transitions (*) and we are done.
\qed

\begin{exa}
In this example we want to construct two components whose asynchronous composition has infinitely many states but weak asynchronous compatibility is decidable.
As a tool we want to apply Thm.~\ref{thm.completenss}. Therefore our components should satisfy properties $\mathcal{X}_A$ and $\mathcal{X}_B$
of Thm.~\ref{thm.completenss}.
Consider the two components \texttt{MA} and \texttt{MB} in Fig.~\ref{fig:weak-asynch}.
We remove the transition  0 $\must{\texttt{failA}}{}{\texttt{MB}}$ 0 with input action \texttt{failA}
from \texttt{MB} which gives us the component \texttt{MB'}.
Now we show that the properties $\mathcal{X}_A$ and $\mathcal{X}_B$ are satisfied
for \texttt{MA} and \texttt{MB'}. 
To check $\mathcal{X}_A$
we must consider the state 2 of \texttt{MA} in which an output is enabled and which is locally reachable in
$\texttt{MA} \stimes \texttt{MB'}^\rhd_{\{\texttt{readyB,failB}\}}$ (since, e.g., (2,0) is reachable in
$\texttt{MA} \stimes \texttt{MB'}^\rhd_{\{\texttt{readyB,failB}\}}$). 
Obviously, the state (2,$\emptyseq$) of $\Omega(\texttt{MA})$ is locally reachable in $\Omega(\texttt{MA}) \stimes \Omega(\texttt{MB'})$.
(For instance,  ((2,$\emptyseq$),(0,$\emptyseq$)) is reachable in $\Omega(\texttt{MA}) \stimes \Omega(\texttt{MB'})$.)
Property $\mathcal{X}_B$ is checked analogously. According to Thm.~\ref{thm.completenss} we can therefore decide
whether \texttt{MA} and \texttt{MB'} are  weakly asynchronously compatible.
In this example, $\texttt{MA} \dashrightarrow \texttt{MB'}^\rhd_{\{\texttt{readyB,failB}\}}$ does \emph{not} hold
since in state (2,0) the component \texttt{MA} can output \texttt{failA} which cannot be accepted by \texttt{MB'}. 
Therefore, by Thm.~\ref{thm.completenss}, we have proved that \texttt{MA} and \texttt{MB'} are \emph{not} weakly asynchronously compatible.
\end{exa}

\section{Deadlock Analysis for Communicating Components}\label{sec:df}

Another property which is important when analysing system behaviours concerns deadlock-freeness.
We are interested here in the analysis of deadlock-freeness for communicating components $A$ and $B$.

\begin{defi}\label{def:df}
Let $A$ and $B$ be two asynchronously composable IOTSes.
\begin{enumerate}
\item
A \emph{deadlock state} of the synchronous system $A \stimes B$ is a state $(s_A,s_B) \in \reach(A \stimes B)$
such that there exists no outgoing transition from  $(s_A,s_B)$ in $A \stimes B$.
If $A \stimes B$ has no deadlock state then it is \emph{synchronously deadlock-free}, denoted by $\df$.
\item
A \emph{deadlock state} of the asynchronous system $A \atimes B$ is a state
$((s_A,q_A),(s_B,q_B)) \in \reach(\Omega(A) \stimes \Omega(B))$
such that there exists no outgoing transition from  $((s_A,q_A),(s_B,q_B))$ in $\Omega(A) \stimes \Omega(B)$.
If $A \atimes B$ has no deadlock state then it is \emph{asynchronously deadlock-free}, denoted by $\dfA$.
\end{enumerate}
\end{defi}

For finite IOTSes $A$ and $B$ synchronous deadlock-freeness is decidable while asynchronous deadlock-freeness is generally undecidable.
In this section we study possibilities for verification of deadlock-freeness for
asynchronous systems. First, we want to point out that deadlock-freeness and (weak) asynchronous compatibility are different properties. None of the two implies the other.

\begin{exa}\label{ex:wac-df}\hfill
\begin{enumerate}
\item
$A \wac B$ does not imply $\dfA$:
We consider two components $A$ and $B$ such that
$\inp_A = \{b\}$, $\out_A = \{a\}$, $\internal_A = \emptyset$ and
$\inp_B = \{a\}$, $\out_B = \{b\}$, $\internal_B = \emptyset$.
The transitions of $A$ and $B$ are shown in Fig.~\ref{fig:wac-df-1}.
Component $A$ is always ready to accept $b$ and $B$ is always ready to accept $a$ but none of the two ever sends a message  to the other.
Hence $A \wac B$ (and also $A \ac B$) holds trivially but, since no message is sent, the initial state of
$\Omega(A) \stimes \Omega(B)$ is a deadlock state.

\begin{figure}
  \centering
 \includegraphics[scale=0.7]{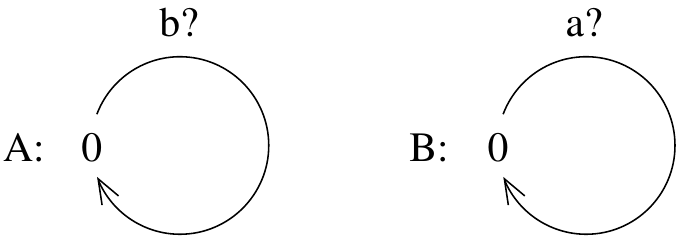}
  \caption{$A \wac B$ but \underline{not} $\dfA$}
  \label{fig:wac-df-1}
\end{figure}

\item
$\dfA$ does not imply $A \wac B$:
Let $A$ and $B$ be two components with the actions defined in part (1) above.
The transitions of $A$ and $B$ are shown in Fig.~\ref{fig:wac-df-2}.
The asynchronous composition $A \atimes B$ is deadlock-free since $A$ puts continuously message $a$ in its
output queue while $B$ does the same with message $b$. Since $A$ (resp.\ $B$) never takes the message addressed to it, the system is not weakly asynchronously compatible (and also not strongly asynchronously compatible).

 \begin{figure}
  \centering
 \includegraphics[scale=0.7]{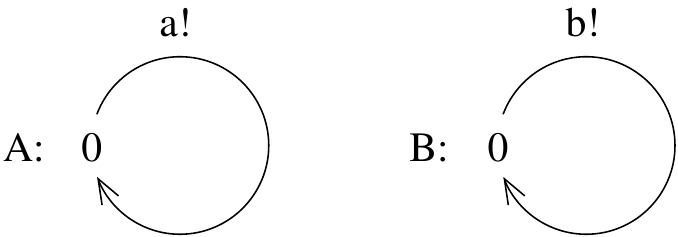}
  \caption{$\dfA$ but \underline{not} $A \wac B$}
  \label{fig:wac-df-2}
\end{figure}
\end{enumerate}
\end{exa} 

In the following of this section we assume that we have already checked that $A$ and $B$ are weakly asynchronously compatible
and that we now want to prove deadlock-freeness of the asynchronous system. The half-duplex property is again useful for this case.
In fact, if the asynchronous system is half-duplex, then deadlock freeness of the asynchronous system is equivalent to deadlock-freeness of the synchronous composition.

\begin{thm}\label{thm:df-half-suplex}
Let $A$ and $B$ be two asynchronously composable and weakly asynchronously compatible IOTSes.
If the asynchronous system $A \atimes B$ is half-duplex, then $\df$ holds if, and, only if $\dfA$ holds.  
\end{thm}

\proof
$\Rightarrow$:
Let $((s_A,q_A),(s_B,q_B))$ be an arbitrary state in $\reach(\Omega(A) \stimes \Omega(B))$.\\
\textit{Case 1:} $q_A \neq \emptyseq$ or $q_B \neq \emptyseq$.
By assumption, $A \wac B$. Hence, any element being in one of the queues will be consumed and therefore
$((s_A,q_A),(s_B,q_B))$ is not a deadlock state of  $A \atimes B$.\\
%W.l.o.g. let $q_A = a_1 \ldots a_n$ with $n \geq 1$.
%Since, by assumption, $A \wac B$, there exist transitions
%$(s_B,q_B) \  \xmust{\internal_{\Omega(B)}}{*}{\Omega(B)} \  (\overline s_B,\overline q_B)$ 
%$\must{a_1}{}{\Omega(B)} (s'_B,\overline q_B)$ .
%Hence we get transitions
%$((s_A,q_A),(s_B,q_B))  \  \xmust{\internal_{\Omega(B)}}{*}{\Omega(A) \stimes \Omega(B)} ((s_A,q_A),(\overline s_B,\overline q_B)) \must{a_1}{}{\Omega(A) \stimes \Omega(B)}$\linebreak
%$((s_A,a_2 \ldots a_n),(s'_B,\overline q_B)).$
%Therefore
%$((s_A,q_A),(s_B,q_B))$ is not a deadlock state of  $A \atimes B$.\\
\textit{Case 2:} Let $q_A = q_B = \emptyseq$.
Since  $A \atimes B$ is half-duplex we then know, by Lem.~\ref{lem:main},
that $A \atimes B$ satisfies property $\mathcal{P}(i)$. Therefore $(s_A,s_B) \in \reach(A \stimes B)$.
Since $\df$ holds, there exists a transition $(s_A,s_B) \xmust{x}{}{A \stimes B} (s'_A,s'_B)$.
If $x$ is a non-shared action of $A$ or of $B$ then this transition is induced by a transition of $A$ or $B$ which
in turn induces a transition of $\Omega(A) \stimes \Omega(B)$ starting in $((s_A,\emptyseq),(s_B,\emptyseq))$. 
In fact, if $x$ is not a shared action of $A$ and $B$ this is clear.
If $x$ is a shared action of $A$ and $B$ there are two cases:
$x \in \out_A \cap \inp_B$ or  $x \in \out_B \cap \inp_A$. W.l.o.g. let $x \in \out_A \cap \inp_B$.
Then $(s_A,s_B) \xmust{x}{}{A \stimes B} (s'_A,s'_B)$ is
induced by transitions $s_A \xmust{x}{}{A} s'_A$ and $s_B \xmust{x}{}{B} s'_B$.
The transition $s_A \xmust{x}{}{A} s'_A$ induces a transition
 $((s_A,\emptyseq),(s_B,\emptyseq)) \xmust{x^\rhd}{}{\Omega(A) \stimes \Omega(B)} ((s'_A,x),(s_B,\emptyseq))$.
Hence, $((s_A,\emptyseq),(s_B,\emptyseq))$ is not a deadlock state of  $A \atimes B$. 
 Thus, in all possible cases $((s_A,q_A),(s_B,q_B))$
is not a deadlock state of  $A \atimes B$ and therefore $\dfA$ holds. 

$\Leftarrow$:
Let $(s_A,s_B)$ be an arbitrary state in $\reach(A \stimes B)$.
By Lem.~\ref{lem:async2synch},
$((s_A,\emptyseq), (s_B,\emptyseq))$ belongs to $\reach(\Omega(A) \stimes \Omega(B))$.
Since $\dfA$ holds, there exists a transition
 $$((s_A,\emptyseq),(s_B,\emptyseq)) \xmust{x}{}{\Omega(A) \stimes \Omega(B)} ((s'_A,q_A),(s'_B,q_B)).$$
If $x$ is an action of $A$ or of $B$ which is not shared between $A$ and $B$,
then this transition is induced by a transition of $A$
or of $B$ which in turn induces a transition of $A \stimes B$ starting in  $(s_A,s_B)$.
Hence,  $(s_A,s_B)$ is not a deadlock state of  $A \stimes B$.
Otherwise there are four cases: (i) $x \in \out_A \cap \inp_B$, (ii) $x \in \out_B \cap \inp_A$,
or (iii) $x$ is of the form $a^\rhd$ with $a \in \out_A \cap \inp_B$
or (iv) $x$ is of the form $b^\rhd$ with $b \in \out_B \cap \inp_A$.
Cases (i) and (ii) are not possible since, e.g., case (i) relies on an input action of $B$
which is not possible since the output queue of $A$ is empty.
For the remaining two cases we consider, w.l.o.g., case (iii).
Then $((s_A,\emptyseq),(s_B,\emptyseq)) \xmust{a^\rhd}{}{\Omega(A) \stimes \Omega(B)} ((s'_A,a),(s_B,\emptyseq))$
is induced by a transition $s_A \xmust{a}{}{A} s'_A$ with $a \in  \out_A \cap \inp_B$.
Since, by assumption, $A \wac B$ holds and $A \atimes B$ is half-duplex,
we know, by Thm.~\ref{thm:async2sync}(2), that  $A \wsc B$ holds. 
Therefore there exist transitions
$ s_B \  \xmust{\internal_B}{*}{B} \  \overline s_B \must{a}{}{B} s'_B$
which induce transitions
$(s_A,s_B) \  \xmust{\internal_B}{*}{A \stimes B} \  (s_A,\overline s_B) \must{a}{}{A\stimes B} (s'_A,s'_B)$.
Hence, $(s_A,s_B)$ is not a deadlock state of  $A \stimes B$. Thus, in all possible cases $(s_A,s_B)$
is not a deadlock state of  $A \stimes B$ and therefore $\df$ holds. 
\qed

The next example shows that Thm.~\ref{thm:df-half-suplex} would not hold without the half-duplex assumption.

\begin{exa}\label{ex:dfA-df}\hfill
\begin{enumerate}
\item
$\dfA$ does not imply $\df$:
Let $A$ and $B$ be two components with actions as in Ex.~\ref{ex:wac-df}.
The transitions of $A$ and $B$ are shown in Fig.~\ref{fig:dfA-df-1}.
$A \atimes B$ is not half-duplex.
Obviously, $A \atimes B$ is deadlock-free but $A \stimes B$ is not.

\begin{figure}
  \centering
 \includegraphics[scale=0.7]{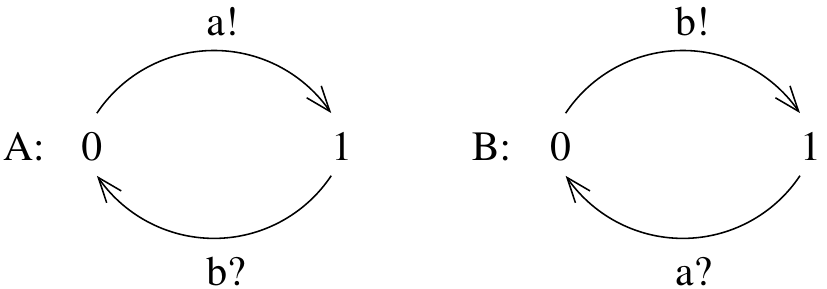}
  \caption{$\dfA$ but \underline{not} $\df$}
  \label{fig:dfA-df-1}
\end{figure}

\item
$\df$ does not imply $\dfA$:
Let $A$ and $B$ be two components with the actions as above but with an additional shared action $x$
being an output action of $A$ and an input action of $B$.
The transitions of $A$ and $B$ are shown in Fig.~\ref{fig:dfA-df-2}.
$A \atimes B$ is not half-duplex.
Obviously, $A \stimes B$ is deadlock-free but $A \atimes B$ is not.

 \begin{figure}
  \centering
 \includegraphics[scale=0.7]{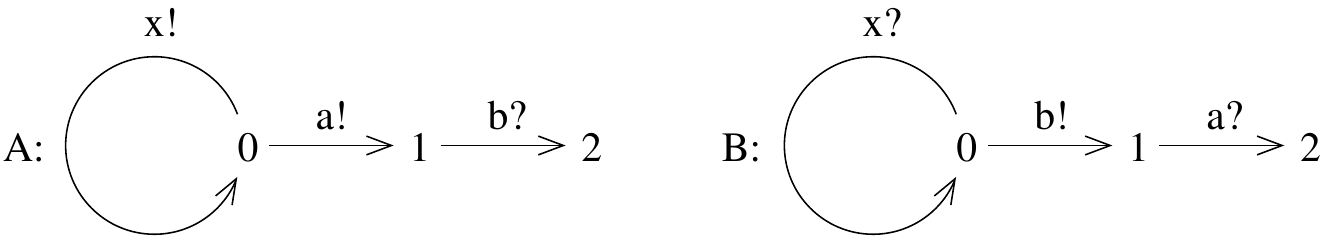}
  \caption{$\df$ but \underline{not} $\dfA$}
  \label{fig:dfA-df-2}
\end{figure}
\end{enumerate}
\end{exa} 

We are now interested in verifying deadlock-freeness in the general case where $A \atimes B$ is not half-duplex.
Similarly to the technique proposed for verifying weak asynchronous compatibility we rely again on a criterion
which uses the synchronous products $A \stimes \renBA$ and $\renAB \stimes B$; see Sect.~\ref{sec:asynch-comp}.

\begin{defi}\label{def:df-1}
Let $A$ and $B$ be two asynchronously composable IOTSes.
$A \stimes \renBA$ is \emph{autonomously deadlock free}
if for each reachable state $(s_A,s_B) \in \reachBA$
there exists a transition $(s_A,s_B) \  \must{a}{}{A \stimes \renBA} (s'_A,s'_B)$ with
$a \notin \inp_A \cap \out_B$. 
Autonomous deadlock-freeness of $\renAB \stimes B$ is defined analogously.
\end{defi}

Note that the condition $a \notin \inp_A \cap \out_B$ is needed in the next theorem
to ensure $\dfA$. Otherwise, if we have a transition
$(s_A,s_B) \  \must{a}{}{A \stimes \renBA} (s'_A,s'_B)$ with
$a \in \inp_A \cap \out_B$, then this would be a free input of $A$
in the composition $A \stimes \renBA$. But in the asynchronous composition
$A \atimes B$, $a$ would be a shared input which can only be performed if
the message $a$ is available in the output queue of $B$. But this may not be the case and therefore
$A \atimes B$ could be in a state in which it cannot continue with $a$ while $A \stimes \renBA$
could continue due to the freeness of $a$.
Therefore deadlock-freeness of $A \stimes \renBA$ or
$\renAB \stimes B$ would not be sufficient and
that's why we have introduced  the stronger version of autonomous deadlock-freeness above.
 
\begin{thm}\label{thm:dfA-general-case}
Let $A$ and $B$ be two asynchronously composable and weakly asynchronously compatible IOTSes.
If $A \stimes \renBA$ \emph{or}  $\renAB \stimes B$ is autonomously deadlock free, then
$\dfA$.
\end{thm}

\proof
As in the proof of Thm.~\ref{thm:df-half-suplex}, direction ``$\Rightarrow$'', the critical cases are states\linebreak
$((s_A,q_A),(s_B,q_B)) \in \reach(\Omega(A) \stimes \Omega(B))$ with $q_A = q_B = \emptyseq$.
(Otherwise, the assumption $A \wac B$ guarantees progress.)
W.l.o.g.\ let $A \stimes \renBA$ be autonomously deadlock free.
Since $ q_A = \emptyseq$, we know, by Lem.~\ref{lem:general-case}, that $(s_A,s_B) \in \reachBA$. 
Then, by assumption, there exists a transition $(s_A,s_B) \  \must{x}{}{A \stimes \renBA} (s'_A,s'_B)$ with
$x \notin \inp_A \cap \out_B$. 
If $x$ is an action of $A$ or of $B$ which is not shared between $A$ and $B$,
then this transition is induced by a transition of $A$ or of $B$ which
in turn induces a transition of $\Omega(A) \stimes \Omega(B)$ starting in $((s_A,\emptyseq),(s_B,\emptyseq))$.
Hence, $((s_A,\emptyseq),(s_B,\emptyseq))$ is not a deadlock state of  $A \atimes B$. 
Otherwise (i) $x \in \out_A \cap \inp_B$ or (ii) $x$ is of the form $b^\rhd$ with $b \in \out_B \cap \inp_A$.
(Note that $x \in \inp_A \cap \out_B$ is not possible due to the assumption.)

\noindent(i): If $x \in \out_A \cap \inp_B$, then
$(s_A,s_B) \xmust{x}{}{A \stimes \renBA} (s'_A,s'_B)$ is
induced by transitions $s_A \xmust{x}{}{A} s'_A$ and $s_B \xmust{x}{}{\renBA} s'_B$.
The transition $s_A \xmust{x}{}{A} s'_A$ induces a transition
$$((s_A,\emptyseq),(s_B,\emptyseq)) \xmust{x^\rhd}{}{\Omega(A) \stimes \Omega(B)} ((s'_A,x),(s_B,\emptyseq)).$$
Hence, $((s_A,\emptyseq),(s_B,\emptyseq))$ is not a deadlock state of  $A \atimes B$.

\noindent(ii): If $x$ is of the form $b^\rhd$ with $b \in \out_B \cap \inp_A$,
then $(s_A,s_B) \xmust{b^\rhd}{}{A \stimes \renBA} (s_A,s'_B)$ is
induced by a transition $s_B \xmust{b^\rhd}{}{\renBA} s'_B$.
Hence, there exists a transition
$$((s_A,\emptyseq),(s_B,\emptyseq)) \xmust{b^\rhd}{}{\Omega(A) \stimes \Omega(B)} ((s_A,\emptyseq),(s'_B,b))$$
and therefore 
$((s_A,\emptyseq),(s_B,\emptyseq))$ is not a deadlock state of  $A \atimes B$. 
In summary, there is no deadlock state of  $A \atimes B$ and therefore $\dfA$ holds. 
\qed

\begin{exa}
Consider the two components \texttt{MA} and \texttt{MB} in Fig.~\ref{fig:weak-asynch}.
We have shown, in Ex.~\ref{exa:general}, that
$\texttt{MA} \wac \texttt{MB}$ holds.
It is easy to check that $\texttt{MA} \stimes \texttt{MB}^\rhd_{\{\texttt{readyB,failB}\}}$
is autonomously deadlock-free, since in any reachable state an action different from
\texttt{readyB,failB} can be executed. Therefore, we can apply Th.~\ref{thm:dfA-general-case}
and get $\textit{df}(\texttt{MA} \atimes \texttt{MB})$. 
\end{exa}

%%%%%%%%%%%%%%%%%%%%%%%%%%%%%%%%%%%%%%%%

\section{Related Work}\label{sec:related-work}
Compatibility notions are mostly considered for synchronous systems, since in this case
compatibility checking is easier manageable and even decidable if the behaviors of local components have finitely many states.
Some approaches use process algebras to study compatibility, like~\cite{DBLP:journals/scp/CanalPT01} using the $\pi$-calculus,
others investigate interface theories 
with binary compatibility relations preserved by refinement, see,
e.g.,
interface automata~\cite{Alfaro2001} or
modal interfaces~\cite{DBLP:journals/fuin/RacletBBCLP11,productlines}.
Others consider n-ary compatibility in multi-component systems like, e.g., team automata~\cite{DBLP:journals/tcs/CarmonaK13}.
A prominent example of multi-component systems with asynchronous communication via unbounded FIFO-buffers are CFSMs~\cite{Brand-Zafiropulo}, for which many problems, like absence of unspecified reception, are undecidable.
%The situation is different, if half-duplex systems of two CFSMs are considered. C{\'{e}}c{\'{e} and Finkel have shown in~\cite{DBLP:journals/iandc/CeceF05} that then the set of reachable configurations is recognizable and several problems, including absence of unspecified receptions, are decidable.
%\cite{DBLP:conf/wsfm/LozesV11} considers the half-duplex property as an appropriate definition for reliable contracts and shows that in this case besides absence of unspecified receptions also boundedness and absence of message orphans is decidable.
Exceptions where decidability is ensured are half-duplex systems consisting of two components; see, e.g.,~\cite{DBLP:journals/iandc/CeceF05} 
and~\cite{DBLP:conf/wsfm/LozesV11},
or systems whose  network topologies are acyclic; see~\cite{DBLP:conf/tacas/TorreMP08}.
Bag structures are typically used for modeling asynchronous communication with Petri nets where the reachability problem,
and therefore many compatibility problems~\cite{DBLP:conf/apn/HaddadHM13}, are decidable.
%see, e.g.~\cite{DBLP:conf/stoc/Mayr81}.
In~\cite{ClementeEtAl2014}  decidable topologies are studied for systems which contain both FIFO and bag channels for communication.

There is, however, not much work on relationships between synchronous and asynchronous compatibility.
Exceptions are approaches based on \emph{synchronizability}~\cite{DBLP:conf/facs2/OuederniSB13}
and on \emph{desynchronizability}~\cite{DBLP:journals/scp/BeoharC14}.
Despite of the different terminologies in both cases the idea is to establish a branching bisimulation between the synchronous and the asynchronous versions of a system with message consumption from buffers considered internal.
Under the assumption of synchronizability~\cite{DBLP:conf/facs2/OuederniSB13} proposes methods to prove compatibility of asynchronously communicating peers by checking synchronous compatibility. The central notion is (synchronous/asynchronous) UR compatibility which corresponds to our weak (synchronous/asynchronous) compatibility plus deadlock-freeness.
Comparing our work to~\cite{DBLP:conf/facs2/OuederniSB13}, obvious differences are that~\cite{DBLP:conf/facs2/OuederniSB13} considers
multi-component systems while we study compatibility for two components only. On the other hand,~\cite{DBLP:conf/facs2/OuederniSB13} considers closed systems while we allow open systems. 
%which can be incrementally extended to larger ones.
Also our method for checking asynchronous compatibility is very different. 
In the first part of our work we rely on half-duplex systems instead of synchronizability
and in the second part we drop any assumptions and investigate powerful and decidable criteria for 
asynchronous compatibility of systems which are neither half-duplex nor synchronizable.

The desynchronization approach in~\cite{DBLP:journals/scp/BeoharC14,beohar-diss} suggests a variant of asynchronous composition which enforces the half-duplex property by blocking outputs to a buffer if there are inputs waiting in the other buffer.~\cite{beohar-diss} shows that for such systems desynchronizability implies freedom of orphans,
which means that buffers can always be emptied and therefore no message loss can occur.
Moreover, conditions for the synchronous system are provided which characterize desynchronizability\footnote{For the definition of a desynchronizable system~\cite{DBLP:journals/scp/BeoharC14,beohar-diss} 
use a variant of branching bisimulation which is sensitive w.r.t. emptyness of buffers.}. Among them is the condition
of well-posedness which coincides with our notion of synchronous strong compatibility.
The results in~\cite{DBLP:journals/scp/BeoharC14,beohar-diss} are established for concrete systems,
 i.e., the underlying components do not have silent transitions. Under this assumption strong and weak synchronous compatibility are the same.
Comparing our work to~\cite{DBLP:journals/scp/BeoharC14,beohar-diss}, obvious differences are that we allow internal transitions of the underlying components which (a) leads to the notion of weak synchronous compatibility
and (b) is also necessary to scale to larger systems where subsystems are asynchronously composed and
therefore introduce silent transitions anyway. Moreover, in the first part of our work we do not enforce half-duplex queues (which are not supported by standard implementation technologies either) but study asynchronous systems which have by themselves the half-duplex property. 
As already mentioned above, in the second part we drop any assumptions such that
we can treat also systems which are not desynchronizable.

Another issue concerns checking the correctness of implementations of reactive
components against their specifications.
Although this is not really a topic of this work, it is still very relevant
that compatibility proved on specification level should 
also hold for (possibly distributed) implementations. Since in practice implementations often use asynchronous message passing,
it would be nice if one could check compatibility for \emph{synchronous} composition of specifications
and infer from this compatibility for \emph{asynchronous} composition of implementations.
A pragmatic solution has been studied in~\cite{DBLP:journals/toplas/YellinS97}
where programming strategies have been proposed to ensure that
implementations of reactive components collaborate correctly if their component protocols are compatible in the sense of strong synchronous compatibility as considered here. 
A formal treatment for implementation correctness using a testing approach has been studied
in~\cite{weiglhofer-diss} and, developed further, in~\cite{DBLP:journals/sosym/NorooziKMW15}.
The testing approach for input-output conformance is motivated by the fact that a formal model of a
concrete implementation may not be available and therefore the implementation can only be checked by running 
a set of test cases against it.
Since  implementations are often only accessible through asynchronous communication channels
the observations obtained by testing an implementation rely on an asynchronous interaction.
On the other hand, test cases and component specifications can be represented by transition systems
with input-output labels
and their interaction could be most easily observed by a synchronous composition of the two.
Then the question at hand is whether observations obtained by synchronous composition
of a set of test cases with a component specification
allow us to infer that an implementation tested in an asynchronous environment conforms to the specification.
\cite{weiglhofer-diss} and~\cite{DBLP:journals/sosym/NorooziKMW15} develop conditions under which such an approach is feasible. Interestingly the conditions are very much related to properties studied in the first part of our paper. For instance, an internal choice input-output labeled transition system
in~\cite{weiglhofer-diss,DBLP:journals/sosym/NorooziKMW15} is an I/O-separated transition system
according to our Def.~\ref{def:io-sep}(1).
% (but not conversely, since I/O-separatedness admits states in which input actions and internal actions are enabled at the same time).
An internal choice test case in
the sense of~\cite{DBLP:journals/sosym/NorooziKMW15} stimulates
an implementation-under-test only when ``quiescence  has been observed'', i.e.,
the implementation does not
send itself an output to the test case in the current state. This is similar to condition (3) in our Lem.~\ref{lem:main}
which expresses that two synchronously composed components cannot reach a state in which each of them has an output enabled.
We have shown in Lem.~\ref{lem:main} that this condition characterizes half-duplex systems.
Similarly, Lemma 4 in~\cite{DBLP:journals/sosym/NorooziKMW15} states when executing an internal choice test case on an implementation behaving as internal choice input-output labeled transition system,
the input and output queues cannot be empty simultaneously, i.e., we get half-duplex communication.
Despite of these technical similarities the goals of our work are quite different since we study
safe communication and not correctness of implementations w.r.t.\ specifications.

Last not least let us point out that the first part of our work is closely related to the study of half-duplex systems by C{\'{e}}c{\'{e} and Finkel~\cite{DBLP:journals/iandc/CeceF05}.
Due to their decidability result concerning unspecified reception (for two communicating CFSMs)
it is not really surprising that we get an effective characterization of asynchronous compatibility
and a way to decide it for components with finitely many states.
A main difference to~\cite{DBLP:journals/iandc/CeceF05} is that we consider
also synchronous systems and relate their compatibility properties to the asynchronous versions.
Moreover, we deal with open systems as well and consider a weak variant
of asynchronous compatibility, which we believe adds much power to the strong version.
The same differentiation applies to~\cite{DBLP:conf/wsfm/LozesV11}.
Finally, as explained above, a significant part of
our work deals with systems which are not necessarily half-duplex
and this was not an issue in~\cite{DBLP:journals/iandc/CeceF05}.

%=======================================================

\section{Conclusion}\label{sec:conclusion}

We have proposed techniques to verify asynchronous compatibility
and deadlock-freeness of communicating components  by using criteria that are based on synchronous composition and hence decidable (if the components are finite state).
We have shown that strong (weak) synchronous and strong (weak) asynchronous
compatibility are equivalent if the asynchronous system is half-duplex.
For non-half duplex systems we have provided decidable conditions
which are sufficient for weak asynchronous compatibility.

The IOTSes used here for modeling component behaviors are special cases
of modal I/O-transition systems~\cite{productlines}, for which synchronous composition and synchronous
compatibility checking is implemented in the MIO Workbench~\cite{tacas2010}, an Eclipse-based verification tool.
Since the verification conditions studied in this paper involve only synchronous compatibility checking, we can use the MIO Workbench for this purpose.

The most important issues for future research concern 
(a) the consideration of other compatibility problems, e.g., that a component waiting for some input will eventually get it~\cite{DBLP:conf/coordination2017},
and (b) 
the extension of our approach
to treat multi-component systems. The latter is a particularly challenging task.
For instance, the results on two component half-duplex systems cannot be directly extended since
systems with $n > 2$ components and pairwise half-duplex communication have the power of Turing machines; see~\cite{DBLP:journals/iandc/CeceF05}.

%Instead of binary compatibility relations we are also interested to adjust the n-ary compatibility predicate
%studied for component assemblies in~\cite{DBLP:journals/acta/HennickerK15} to the asynchronous case.

%For closed systems of finite IOTSes without internal actions the behavior described by their asynchronous composition coincides (up to naming conventions)
%with the operational model of communicating finite state machines (CFSMs) in~\cite{Brand-Zafiropulo}.
%In~\cite{Brand-Zafiropulo} a condition is formulated which requires that in a system of CFSMs executable receptions should be \emph{specified},
%which is just the strong version of the asynchronous compatibility notion used here.
%It is, however, important to notice that the weak asynchronous compatibility notion adds much flexibility
%since it allows a component to put first its own issued messages in its output queue (which are internal enqueue actions)
%before it takes available inputs already waiting in the other queue. An example for this situation will be provided later
%in Ex.~\ref{ex:wac}.

\section*{Acknowledgment}
  \noindent 
We are grateful to Alexander Knapp for his suggestion
to use output queues (instead of input queues) for the
formalization of asynchronous compatibility.
Moreover, we would like to thank anonymous reviewers for many useful comments and suggestions.

\appendix
\section{Transitions of \texorpdfstring{$\Omega(A) \stimes \Omega(B)$}{Omega(A) x Omega(B)}}\label{sec:appendix-A}
%
%\noindent\textit{\bf Transitions of $\Omega(A) \stimes \Omega(B)$:}

\begin{itemize}
  \setlength{\itemsep}{1em}
\item If $a \in \inp_{\Omega(A) \stimes \Omega(B)}$:
\begin{itemize}
\item
$a \in \inp_A \smallsetminus \out_B$ and $s_A  \must{a}{}{A} s'_A$\\
then $((s_A,q_A),(s_B,q_B)) \must{a}{}{\Omega(A) \stimes \Omega(B)} ((s'_A,q_A),(s_B,q_B))$, 
\item
$a \in \inp_B \smallsetminus \out_A$ and $s_B  \must{a}{}{B} s'_B$\\
then $((s_A,q_A),(s_B,q_B)) \must{a}{}{\Omega(A) \stimes \Omega(B)} ((s_A,q_A),(s'_B,q_B))$.
\end{itemize}

\item If $a \in \out_{\Omega(A) \stimes \Omega(B)}$:
\begin{itemize}
\item
$a \in \out_A \smallsetminus \inp_B$ and $s_A  \must{a}{}{A} s'_A$\\
then $((s_A,q_A),(s_B,q_B)) \must{a}{}{\Omega(A) \stimes \Omega(B)} ((s'_A,q_A),(s_B,q_B))$, 
\item
$a \in \out_B \smallsetminus \inp_A$ and $s_B  \must{a}{}{B} s'_B$\\
then $((s_A,q_A),(s_B,q_B)) \must{a}{}{\Omega(A) \stimes \Omega(B)} ((s_A,q_A),(s'_B,q_B))$.
\end{itemize}

\item If $a \in  \internal_{A \stimes B} = \internal_A \cup \internal_B \cup (\out_A \cap \inp_B) \cup (\out_B \cap \inp_A)$:%\subseteq \internal_{\Omega(A) \stimes \Omega(B)}$:
\begin{itemize}
\item
$a \in \internal_A$ and $s_A  \must{a}{}{A} s'_A$\\
then $((s_A,q_A),(s_B,q_B)) \must{a}{}{\Omega(A) \stimes \Omega(B)} ((s'_A,q_A),(s_B,q_B))$, 
\item
$a \in \internal_B$ and $s_B  \must{a}{}{B} s'_B$\\
then $((s_A,q_A),(s_B,q_B)) \must{a}{}{\Omega(A) \stimes \Omega(B)} ((s_A,q_A),(s'_B,q_B))$. 
\item
$a \in \out_A \cap \inp_B$ (hence $a \in \out_{Q_{\outAB}}$) and $s_B  \must{a}{}{B} s'_B$\\
then $((s_A,a q_A),(s_B,q_B)) \must{a}{}{\Omega(A) \stimes \Omega(B)} ((s_A,q_A),(s'_B,q_B))$, 
\item
$a \in \out_B \cap \inp_A$ (hence $a \in \out_{Q_{\outBA}}$) and $s_A  \must{a}{}{A} s'_A$\\
then $((s_A,q_A),(s_B,a q_B)) \must{a}{}{\Omega(A) \stimes \Omega(B)} ((s'_A,q_A),(s_B,q_B))$.
\end{itemize}

\item If $a^\rhd \in  \shared(A,B)^\rhd =  (\out_A \cap \inp_B)^\rhd \cup  (\out_B \cap \inp_A)^\rhd$:%\subseteq \internal_{\Omega(A) \stimes \Omega(B)}$:

\begin{itemize}
\item
$a^\rhd \in (\out_A \cap \inp_B)^\rhd$ (hence $a^\rhd \in \inp_{Q_{\outAB}})$
and $s_A  \must{a}{}{A} s'_A$\\%M_A^\rhd)$\\
then $(s_A,q_A) \must{a^\rhd}{}{\Omega(A)} (s'_A,q_Aa) $ and\\
then $((s_A,q_A),(s_B,q_B)) \must{a^\rhd}{}{\Omega(A) \stimes \Omega(B)} ((s'_A,q_A a),(s_B,q_B))$,

\item
$a^\rhd \in (\out_B \cap \inp_A)^\rhd$ (hence $a \in \inp_{Q_{\outBA}})$
and $s_B  \must{a}{}{B} s'_B$\\%M_B^\rhd)$\\
then $(s_B,q_B) \must{a^\rhd}{}{\Omega(B)} (s'_B,q_Ba) $ and\\
then $((s_A,q_A),(s_B,q_B)) \must{a^\rhd}{}{\Omega(A) \stimes \Omega(B)} ((s_A,q_A),(s'_B,q_B a))$.
\end{itemize}
\end{itemize}

\section{Proofs}\label{sec:appendix-B}

\noindent\textit{\bf Proof of Lemma~\ref{lem:main}:}\\
It remains to prove (3) $\Rightarrow$ (1):
We have to show that for each reachable state in $\reach(\Omega(A) \stimes \Omega(B))$
one of the conditions (i), (ii), or (iii) in the definition of property $\mathcal{P}$ is valid.
The initial state $((\start_A,\emptyseq),(\start_B,\emptyseq))$ satisfies (i).
Now assume given an arbitrary transition
$$(*)\hspace{5mm} ((s_A,q_A),(s_B,q_B)) \must{a}{}{\Omega(A) \stimes \Omega(B)} ((s'_A,q'_A),(s'_B,q'_B))$$
with reachable state $((s_A,q_A),(s_B,q_B))$. It is sufficient to show that for any kind of
action $a \in \act_{\Omega(A) \stimes \Omega(B)}$,
if $((s_A,q_A),(s_B,q_B))$ satisfies one of the conditions (i), (ii), or (iii) then
$((s'_A,q'_A),(s'_B,q'_B))$ satisfies (i), (ii), or (iii). The proof is done by case distinction
on the form of the action $a$. %taking into account the case distinction after Def.~\ref{def:asynch-composition}.

\noindent$\mathit{\bf Case~1:}$
In this case we consider actions $a \in \act_A \smallsetminus \shared(A,B)$
which can freely occur in $A$, i.e.\ without involving $B$ or the output queue of $A$.  
This covers the cases 
$a \in \inp_A \smallsetminus \out_B$, $a \in \out_A \smallsetminus \inp_B$, and $a \in \internal_A$.
In all these cases the transition (*) has the form
$$((s_A,q_A),(s_B,q_B)) \must{a}{}{\Omega(A) \stimes \Omega(B)} ((s'_A,q_A),(s_B,q_B))$$
and is induced by a transition $s_A  \must{a}{}{A} s'_A$. It is trivial that, if $((s_A,q_A),(s_B,q_B))$ satisfies (i) ((iii) resp.), then $((s'_A,q_A),(s_B,q_B))$ satisfies (i) ((iii) resp.). 
If the state
$((s_A,q_A),(s_B,q_B))$ satisfies (ii), then $q_A = a_1 \ldots a_m \neq \emptyseq$ and $q_B = \emptyseq$ and there exists $r_A \in \states_A$ such that: $(r_A,s_B) \in \reach(A \stimes B)$ and 
$r_A \Must{a_1}{}{A} \ldots \Must{a_m}{}{A} s_A$.
Since $\Must{a_m}{}{A}$ can involve, besides $a_m$, arbitrary free actions of $A$ and $s_A  \must{a}{}{A} s'_A$
is such a free action, we obtain $r_A \Must{a_1}{}{A} \ldots \Must{a_m}{}{A} s'_A$.
Thus $((s'_A,q_A),(s_B,q_B))$ satisfies (ii).

\noindent$\mathit{\bf Case~2:}$
In this case we consider actions $a \in \act_B \smallsetminus \shared(A,B)$
which can freely occur in $B$, i.e.\ without involving $A$ or the output queue of $B$.
This case is proved analogously to case 1. 

\noindent$\mathit{\bf Case~3:}$
$a \in \out_A \cap \inp_B$. Then the transition (*) has the form
$$((s_A,a q_A),(s_B,q_B)) \must{a}{}{\Omega(A) \stimes \Omega(B)} ((s_A,q_A),(s'_B,q_B))$$
and is induced by a transition $s_B  \must{a}{}{B} s'_B$. 
In this case $((s_A,a q_A),(s_B,q_B))$ can only satisfy (ii) such that:
$q_A = a a_2 \ldots a_m \neq \emptyseq$ and $q_B = \emptyseq$ and there exists $r_A \in \states_A$ such that
$(r_A,s_B) \in \reach(A \stimes B)$ and 
$r_A \Must{a}{}{A} \overline r_A \Must{a_2}{} \ldots \Must{a_m}{}{A} s_A$.
Thereby $r_A \Must{a}{}{A} \overline r_A$ is of the form
$r_A \xmust{F_A}{*}{A} s \must{a}{}{A} s' \xmust{F_A}{*}{A}  \overline r_A$.
Since $F_A$ involves only free actions of $A$ (not shared with $B$),
and since $(r_A,s_B) \in \reach(A \stimes B)$ we have that $(s,s_B) \in \reach(A \stimes B)$.
Now the two transitions $s \must{a}{}{A} s'$ and $s_B  \must{a}{}{B} s'_B$ synchronize and reach
$(s',s'_B) \in \reach(A \stimes B)$. Obviously, $s' \Must{a_2}{} \ldots \Must{a_m}{}{A} s_A$.
If $m=0$ then $q_A = q_B = \emptyseq$ and $s' \xmust{F_A}{*}{A}  s_A$. Thus $(s_A,s'_B) \in \reach(A \stimes B)$ and condition (i) is valid for
$((s_A,q_A),(s'_B,q_B))$. 
Otherwise, since  $(s',s'_B) \in \reach(A \stimes B)$
and $s' \Must{a_2}{} \ldots \Must{a_m}{}{A} s_A$, condition (ii) holds for $((s_A,q_A),(s'_B,q_B))$.

\noindent$\mathit{\bf Case~4:}$
$a \in \out_B \cap \inp_A$.
This case is analogous to case 3. 

\noindent$\mathit{\bf Case~5:}$ $a^\rhd \in (\out_A \cap \inp_B)^\rhd$. Then the transition (*) has the form
$$((s_A,q_A),(s_B,q_B)) \must{a^\rhd}{}{\Omega(A) \stimes \Omega(B)} ((s'_A,q_A a),(s_B,q_B))$$
and is induced by a transition $s_A  \must{a}{}{A} s'_A$  with $a \in \out_A \cap \inp_B$.

If $((s_A,q_A),(s_B,q_B))$ satisfies condition (i), then
$q_A = q_B = \emptyseq$ and $(s_A,s_B) \in \reach(A \stimes B)$. 
Since  $s_A  \must{a}{}{A} s'_A$ we have $s_A  \Must{a}{}{A} s'_A$. 
Thus, taking $r_A = s_A$ condition (ii) is satisfied for $((s'_A,q_A a),(s_B,q_B))$.

If $((s_A,q_A),(s_B,q_B))$ satisfies condition (ii), then
$q_A = a_1 \ldots a_m \neq \emptyseq$ and $q_B = \emptyseq$ and there exists $r_A \in \states_A$ such that:
$(r_A,s_B) \in \reach(A \stimes B)$ and 
$r_A \Must{a_1}{}{A} \ldots \Must{a_m}{}{A} s_A$.
Since $s_A  \must{a}{}{A} s'_A$ we get a sequence
$r_A \Must{a_1}{}{A} \ldots \Must{a_m}{}{A} s_A \Must{a}{}{A} s'_A$.
Thus $((s'_A,q_A a),(s_B,q_B))$ satisfies condition (ii).

If $((s_A,q_A),(s_B,q_B))$ would satisfy condition (iii), then
$q_A = \emptyseq$ and $q_B = b_1 \ldots b_m \neq \emptyseq$ and there exists $r_B \in \states_B$ such that:
$(s_A,r_B) \in \reach(A \stimes B)$ and 
$r_B \Must{b_1}{}{B} \overline r_B \Must{b_2}{} \ldots$ $\Must{b_m}{}{B} s_B$.
Here $r_B \Must{b_1}{}{B} \overline r_B$ has the form
$r_B \xmust{F_B}{*}{B} s \must{b_1}{}{B} s' \xmust{F_B}{*}{B}  \overline r_B$.
Since $F_B$ involves only free actions of $B$ (not shared with $A$),
and since $(s_A,r_B) \in \reach(A \stimes B)$ we get $(s_A,s) \in \reach(A \stimes B)$.
Now we have two transitions $s_A  \must{a}{}{A} s'_A$ with $a \in (\out_A \cap \inp_B)$
and $s \must{b_1}{}{B} s'$ with $b_1 \in (\out_B \cap \inp_A)$ which contradicts the assumption (3).
Hence $((s_A,q_A),(s_B,q_B))$ cannot satisfy condition (iii).

\noindent$\mathit{\bf Case~6:}$
$a^\rhd \in (\out_B \cap \inp_A)^\rhd$.
This case is analogous to case 5.
\qed

%%=============================================================
\noindent\textit{\bf Proof of Lemma~\ref{lem:async2synch}:}\\
The proof is by induction on the length of the derivation of $(s_A,s_B) \in \reach(A \stimes B)$.
For the initial state $(\start_A,\start_B)$ of $A \stimes B$ we have $((\start_A,\emptyseq),(\start_B,\emptyseq)) \in \reach(\Omega(A) \stimes \Omega(B))$.
For the induction step it is enough to show that whenever a state $(s_A,s_B) \in \reach(A \stimes B)$ satisfies
$((s_A,\emptyseq), (s_B,\emptyseq)) \in \reach(\Omega(A) \stimes \Omega(B))$, then for any possible transition
$$(*)\hspace{5mm}(s_A,s_B) \must{a}{}{A \stimes B} (s'_A,s'_B)$$
the successor state $(s'_A,s'_B)$ satisfies $((s'_A,\emptyseq), (s'_B,\emptyseq)) \in \reach(\Omega(A) \stimes \Omega(B))$.
The proof is done by case distinction on the form of the action $a$.

\noindent$\mathit{\bf Case~1:}$ $a \in \act_A\smallsetminus\shared(A,B)$. Then the transition (*) has the form
\[(s_A,s_B) \must{a}{}{A \stimes B} (s'_A,s_B)\] and is induced by a transition $s_A \must{a}{}{A} s'_A$.
We assume
$((s_A,\emptyseq), (s_B,\emptyseq)) \in \reach(\Omega(A) \stimes \Omega(B))$.
Since $a$ is not shared with $B$, the transition $s_A \must{a}{}{A} s'_A$ induces a transition
$((s_A,\emptyseq),(s_B,\emptyseq)) \must{a}{}{\Omega(A) \stimes \Omega(B)} ((s'_A,\emptyseq),(s_B,\emptyseq))$.
Since
$((s'_A,\emptyseq),(s_B,\emptyseq))  \in \reach(\Omega(A) \stimes \Omega(B))$,
$(s'_A,s_B)$ satisfies the desired property.

\noindent$\mathit{\bf Case~2:}$ $a \in \act_B\smallsetminus\shared(A,B)$.
The proof is symmetric to Case 1. 

\noindent$\mathit{\bf Case~3:}$ $a \in \out_A \cap \inp_B$. Then the transition (*)
is induced by two transition $s_A \must{a}{}{A} s'_A$ with $a \in \out_A$
and $s_B \must{a}{}{B} s'_B$ with $a \in \inp_B$. 
We assume $((s_A,\emptyseq), (s_B,\emptyseq)) \in \reach(\Omega(A) \stimes \Omega(B))$.
Since $a \in \out_A \cap \inp_B$, we get a transition
$$((s_A,\emptyseq),(s_B,\emptyseq)) \must{a^\rhd}{}{\Omega(A) \stimes \Omega(B)} ((s'_A,a),(s_B,\emptyseq))$$
with enqueue action $a^\rhd$.
On the other hand, the transition $s_B \must{a}{}{B} s'_B$ gives rise to a transition
$((s'_A,a),(s_B,\emptyseq)) \must{a}{}{\Omega(A) \stimes \Omega(B)} ((s'_A,\emptyseq),(s'_B,\emptyseq))$
 with dequeue action $a$.
Since $((s'_A,\emptyseq),(s'_B,\emptyseq))  \in$ $\reach(\Omega(A) \stimes \Omega(B))$,
$(s'_A,s'_B)$ satisfies the desired property.

\noindent$\mathit{\bf Case~4:}$ $a \in \out_B \cap \inp_A$.
The proof is symmetric to Case 3. 
\qed

%%=============================================================

%%=========================================================
\noindent\textit{\bf Proof of Lemma~\ref{lem:general-case}:}\\
The initial state $((\start_A,\emptyseq),(\start_B,\emptyseq))$ satisfies $\mathcal{Q}_A$ and  $\mathcal{Q}_B$.
Then we consider transitions
$$(*)\hspace{5mm} ((s_A,q_A),(s_B,q_B)) \must{a}{}{\Omega(A) \stimes \Omega(B)} ((s'_A,q'_A),(s'_B,q'_B))$$
and show that if $ ((s_A,q_A),(s_B,q_B))$ satisfies  $\mathcal{Q}_A$ and $\mathcal{Q}_B$ then
$((s'_A,q'_A),(s'_B,q'_B))$ satisfies  $\mathcal{Q}_A$ and  $\mathcal{Q}_B$. The proof is performed by case distinction
on the form of the action $a$. %taking into account the case distinction after Def.~\ref{def:asynch-composition}.
Then the result follows by induction on the length of the sequence of transitions to reach an arbitrary state
$((s_A,q_A),(s_B,q_B)) \in \reach(\Omega(A) \stimes \Omega(B))$. 
In the following we show that property $\mathcal{Q}_A$ is preserved by transitions (*).
For $\mathcal{Q}_B$ the proof is completely analogous.

\noindent$\mathit{\bf Case~1:}$
In this case we consider actions $a \in \act_A \smallsetminus \shared(A,B)$
which can freely occur in $A$, i.e.\ without involving $B$ or the output queue of $A$.  
This covers the cases 
$a \in \inp_A \smallsetminus \out_B$, $a \in \out_A \smallsetminus \inp_B$, and $a \in \internal_A$.
In all these cases the transition (*) has the form
$$((s_A,q_A),(s_B,q_B)) \must{a}{}{\Omega(A) \stimes \Omega(B)} ((s'_A,q_A),(s_B,q_B))$$
and is induced by a transition $s_A  \must{a}{}{A} s'_A$.
If $((s_A,q_A),(s_B,q_B))$ satisfies (i), then $q_A = \emptyseq$ and $(s_A,s_B) \in \reachBA$.
Since $a \in \act_A \smallsetminus \shared(A,B)$ and $A$ and $B$ are asynchronously composable,
$a \in \act_A \smallsetminus \shared(A,\renBA)$.
Hence, since $s_A  \must{a}{}{A} s'_A$, also $(s'_A,s_B) \in \reachBA$
and therefore $((s'_A,q_A),(s_B,q_B))$ satisfies (i).

If
$((s_A,q_A),(s_B,q_B))$ satisfies (ii), then $q_A = a_1 \ldots a_m \neq \emptyseq$ and there exists $r_A \in \states_A$ such that:
$(r_A,s_B) \in \reachBA$ and 
$r_A \xtRarrA{a_1} \ldots \xtRarrA{a_m} s_A$. 
Since $\xtRarrA{a_m}$ can involve, besides $a_m$, arbitrary actions of $A$ which are not in $\out_A \cap \inp_B$ 
and $s_A  \must{a}{}{A} s'_A$
is such a free action, we obtain $r_A \xtRarrA{a_1} \ldots \xtRarrA{a_m} s'_A$.
Thus $((s'_A,q_A),(s_B,q_B))$ satisfies (ii).

\noindent$\mathit{\bf Case~2:}$
In this case we consider actions $b \in \act_B \smallsetminus \shared(A,B)$
which can freely occur in $B$, i.e.\ without involving $A$ or the output queue of $B$.  
This covers the cases 
$b \in \inp_B \smallsetminus \out_A$, $b \in \out_B \smallsetminus \inp_A$, and $b \in \internal_B$.
In all these cases the transition (*) has the form
$$((s_A,q_A),(s_B,q_B)) \must{b}{}{\Omega(A) \stimes \Omega(B)} ((s_A,q_A),(s'_B,q_B))$$
and is induced by a transition $s_B  \must{b}{}{B} s'_B$.
If $((s_A,q_A),(s_B,q_B))$ satisfies (i), then $q_A = \emptyseq$ and $(s_A,s_B) \in \reachBA$.
Since $b \in \act_B \smallsetminus \shared(A,B)$ and $A$ and $B$ are asynchronously composable,
$b \in \act_B \smallsetminus \shared(A,\renBA)$.
Hence, since  $s_B  \must{b}{}{B} s'_B$, also
$s_B  \must{b}{}{\renBA} s'_B$ and $(s_A,s'_B) \in \reachBA$. Therefore
$((s_A,q_A),(s'_B,q_B))$ satisfies (i).

If
$((s_A,q_A),(s_B,q_B))$ satisfies (ii), then $q_A = a_1 \ldots a_m \neq \emptyseq$ and there exists $r_A \in \states_A$ such that:
$(r_A,s_B) \in \reachBA$ and 
$r_A \xtRarrA{a_1} \ldots \xtRarrA{a_m} s_A$.
Since $s_B  \must{b}{}{B} s'_B$ involves only a free action of $B$ and hence of $\renBA$,
$(r_A,s'_B) \in \reachBA$ and therefore $((s_A,q_A),(s'_B,q_B))$ satisfies (ii).

\noindent$\mathit{\bf Case~3:}$
$a \in \out_A \cap \inp_B$. Then the transition (*) has the form
$$((s_A,a q_A),(s_B,q_B)) \must{a}{}{\Omega(A) \stimes \Omega(B)} ((s_A,q_A),(s'_B,q_B))$$
and is induced by a transition $s_B  \must{a}{}{B} s'_B$. 
In this case $((s_A,a q_A),(s_B,q_B))$ can only satisfy (ii) such that:
$q_A = a a_2 \ldots a_m \neq \emptyseq$ and  there exists $r_A \in \states_A$ such that:
$(r_A,s_B) \in \reachBA$ and 
$r_A \xtRarrA{a} \overline r_A \xtRarrA{a_2} \ldots \xtRarrA{a_m} s_A$.
Thereby $r_A  \xtRarrA{a} \overline r_A$ is of the form
$r_A \xmust{Y_A}{*}{A} s \must{a}{}{A} s' \xmust{Y_A}{*}{A}  \overline r_A$.
Since $Y_A$ involves arbitrary actions of $A$ but no action in $\out_A \cap \inp_B$,
and since $(r_A,s_B) \in \reachBA$ 
we have that $(s,s_B) \in \reachBA$.\footnote{Note that the shared
actions of $A$ and $\renBA$ are $\out_A \cap \inp_B$.}
Now the transition $s \must{a}{}{A} s'$ can synchronize with $s_B  \must{a}{}{B} s'_B$ 
and therefore also with $s_B  \must{a}{}{\renBA} s'_B$.
Thus $(s',s'_B) \in \reachBA$. 
Obviously, $s'  \xtRarrA{a_2} \ldots  \xtRarrA{a_m} s_A$.
If $m<2$ then $q_A = \emptyseq$ and $s' \xmust{Y_A}{*}{A}  s_A$.
Thus $(s_A,s'_B) \in \reachBA$ and condition (i) is valid for
$((s_A,q_A),(s'_B,q_B))$. Otherwise, since  $(s',s'_B) \in \reachBA$
and $s'  \xtRarrA{a_2} \ldots  \xtRarrA{a_m} s_A$, condition (ii) holds for $((s_A,q_A),(s'_B,q_B))$.

\noindent$\mathit{\bf Case~4:}$
$a \in \out_B \cap \inp_A$.
Then the transition (*) has the form
$$((s_A,q_A),(s_B,a q_B)) \must{a}{}{\Omega(A) \stimes \Omega(B)} ((s'_A,q_A),(s_B,q_B))$$
and is induced by a transition $s_A  \must{a}{}{A} s'_A$.
If $((s_A,q_A),(s_B, a q_B))$ satisfies (i), then $q_A = \emptyseq$ and $(s_A,s_B) \in \reachBA$.
Since  $s_A  \must{a}{}{A} s'_A$ and $a$ is not a shared action of $A$ and $\renBA$,
since $\outBA =  \out_B \cap \inp_A$ has been renamed to $\renBA$,
also $(s'_A,s_B) \in \reachBA$
and therefore $ ((s'_A,q_A),(s_B,q_B))$ satisfies (i).

If
$((s_A,q_A),(s_B,a q_B))$ satisfies (ii), then $q_A = a_1 \ldots a_m \neq \emptyseq$ and there exists $r_A \in \states_A$ such that:
$(r_A,s_B) \in \reachBA$ and 
$r_A \xtRarrA{a_1} \ldots \xtRarrA{a_m} s_A$.
Since $s_A  \must{a}{}{A} s'_A$ and $a$ is not in $\out_A \cap \inp_B$
we get $r_A \xtRarrA{a_1} \ldots \xtRarrA{a_m} s'_A$.
Thus $((s'_A,q_A),(s_B,q_B))$ satisfies (ii).

\noindent$\mathit{\bf Case~5:}$ $a^\rhd \in (\out_A \cap \inp_B)^\rhd$. Then the transition (*) has the form
$$((s_A,q_A),(s_B,q_B)) \must{a^\rhd}{}{\Omega(A) \stimes \Omega(B)} ((s'_A,q_A a),(s_B,q_B))$$
and is induced by a transition $s_A  \must{a}{}{A} s'_A$ with $a \in \out_A \cap \inp_B$.

If $((s_A,q_A),(s_B, q_B))$ satisfies (i), then $q_A = \emptyseq$ and $(s_A,s_B) \in \reachBA$.
Since  $s_A  \must{a}{}{A} s'_A$ we have $s_A  \xtRarrA{a} s'_A$. 
Thus, taking $r_A = s_A$ condition (ii) is satisfied for $((s'_A,q_A a),(s_B,q_B))$.

If
$((s_A,q_A),(s_B,q_B))$ satisfies (ii), then $q_A = a_1 \ldots a_m \neq \emptyseq$ and there exists $r_A \in \states_A$ such that:
$(r_A,s_B) \in \reachBA$ and 
$r_A \xtRarrA{a_1} \ldots \xtRarrA{a_m} s_A$.
Since $s_A  \must{a}{}{A} s'_A$ we get a sequence
$r_A  \xtRarrA{a_1} \ldots  \xtRarrA{a_m} s_A  \xtRarrA{a} s'_A$.
Thus $((s'_A,q_A a),(s_B,q_B))$ satisfies condition (ii).

\noindent$\mathit{\bf Case~6:}$
$b^\rhd \in (\out_B \cap \inp_A)^\rhd = \outBA^\rhd$. Then the transition (*) has the form
$$((s_A,q_A),(s_B,q_B)) \must{b^\rhd}{}{\Omega(A) \stimes \Omega(B)} ((s_A,q_A),(s'_B,q_B b))$$
and is induced by a transition $s_B  \must{b}{}{B} s'_B$ with $b \in \out_B \cap \inp_A = \outBA$.

If $((s_A,q_A),(s_B,q_B))$ satisfies (i), then $q_A = \emptyseq$ and $(s_A,s_B) \in \reachBA$.
Since $s_B  \must{b}{}{B} s'_B$ we have $s_B  \must{b^\rhd}{}{\renBA} s'_B$.
Moreover, since $A$ and $B$ are asynchronously composable, $b^\rhd$ is not a shared action with $A$. 
Hence $(s_A,s'_B) \in \reachBA$. Thus $((s_A,q_A),(s'_B,q_B b))$ satisfies (i). 

If $((s_A,q_A),(s_B,q_B))$ satisfies (ii), then $q_A = a_1 \ldots a_m \neq \emptyseq$ and there exists $r_A \in \states_A$ such that:
$(r_A,s_B) \in \reachBA$ and 
$r_A \xtRarrA{a_1} \ldots \xtRarrA{a_m} s_A$.
Since $s_B  \must{b}{}{B} s'_B$ we have $s_B  \must{b^\rhd}{}{\renBA} s'_B$
and since $b^\rhd$ is not a shared action with $A$ we get
$(r_A,s'_B) \in \reachBA$.
Thus $((s_A,q_A),(s'_B,q_B b))$ satisfies (ii). 
\qed


\begin{thebibliography}{50}

\bibitem{Alfaro2001}
  Luca de Alfaro and Thomas A. Henzinger.
  \newblock Interface Automata.
  \newblock  Proc. {\em 9th ACM SIGSOFT Ann. Symp. Foundations of Software Engineering (FSE'01)}, 109--120.
  \newblock ACM Press, 2001.

%\bibitem{DBLP:conf/popl/BasuBO12}
% Samik Basu,
%               Tevfik Bultan, and
%               Meriem Ouederni.
%   \newblock Deciding Choreography Realizability.
%    \newblock Proc. {\em {ACM} {SIGPLAN-SIGACT} Symposium on Principles
%               of Programming Languages, {POPL'12}},
%    \newblock 191--202,
%  \newblock {ACM}, 2012.
%  
%\bibitem{DBLP:conf/vmcai/BasuBO12}
% Samik Basu,
%               Tevfik Bultan, and
%               Meriem Ouederni.
%   \newblock Synchronizability for Verification of Asynchronously Communicating
%               Systems.
%   \newblock Proc.{\em Verification, Model Checking, and Abstract Interpretation {VMCAI'12}},
%\newblock Lecture Notes in Computer Science 7148, 56--71.
%\newblock Springer, 2012.

\bibitem{DBLP:journals/corr/abs-1101-4731}
  Sebastian S. Bauer,
               Rolf Hennicker, and
               Stephan Janisch.
  \newblock Interface Theories for (A)synchro\-nously Communicating Modal
               {I/O}-Transition Systems.
  \newblock Proc. {\em Foundations for Interface Technologies, {FIT'10}},
  \newblock EPTCS 46, 1--8, 2010.

\bibitem{tacas2010}
  Sebastian S. Bauer, Philip Mayer, Andreas Schroeder, and Rolf
	Hennicker.
 \newblock On Weak Modal Compatibility, Refinement, and the {MIO Workbench}.
  \newblock Proc. {\em 16th Int. Conf. Tools and Algorithms for the Construction and Analysis of Systems (TACAS'10)},
\newblock Lecture Notes in Computer Science 6015, 175--189.
\newblock Springer, 2010.
 

\bibitem{Brand-Zafiropulo}
  Daniel Brand and
               Pitro Zafiropulo.
  \newblock On Communicating Finite-State Machines.
   \newblock {\em J. ACM}, 30(2), 323--342, 1983.

\bibitem{DBLP:conf/coordination2017}
Maurice~H.~ter~Beek,
Josep Carmona,
Rolf Hennicker, and
Jetty Kleijn.
\newblock Communication Requirements for Team Automata.
\newblock Proc. {\em 19th IFIP Int. Conf. on Coordination Models and Languages (COORDINATION'17)},
\newblock Lecture Notes in Computer Science.
\newblock Springer, to appear 2017.

\bibitem{DBLP:journals/scp/BeoharC14}
  Harsh Beohar, and
  Pieter J. L. Cuijpers.
\newblock Avoiding Diamonds in Desynchronisation.
\newblock {\em Sci. Comput. Program.}, 91, 45-69, 2014.

\bibitem{beohar-diss}
  Harsh Beohar.
\newblock Refinement of Communication and States in Models of Embedded Systems.
\newblock Faculty of Mathematics and Computer Science, Technische Universiteit Eindhoven, 2013.

\bibitem{DBLP:journals/scp/CanalPT01}
  Carlos Canal,
               Ernesto Pimentel, and
               Jos{\'{e}} M. Troya.
  \newblock Compatibility and Inheritance in Software Architectures.
  \newblock {\em Sci. Comput. Program.}, 41(2), 105--138, 2001.

%\bibitem{DBLP:conf/fase/CanalS15}
% Carlos Canal and
%               Gwen Sala{\"{u}}n},
%  \newblock Model-Based Adaptation of Software Communicating via {FIFO} Buffers.
%  \newblock Proc. {\em Fundamental Approaches to Software Engineering - 18th Int.
%               Conf. {FASE'15}},
%\newblock Lecture Notes in Computer Science 9033, 252--266.
%\newblock Springer, 2015.

\bibitem{DBLP:journals/tcs/CarmonaK13}
 Josep Carmona and
               Jetty Kleijn.
 \newblock Compatibility in a Multi-component Environment.
  \newblock {\em Theor. Comput. Sci.}, 484, 1--15, 2013.

\bibitem{DBLP:journals/iandc/CeceF05}
  G{\'{e}}rard C{\'{e}}c{\'{e}} and
               Alain Finkel.
  \newblock Verification of Programs with Half-duplex Communication.
  \newblock {\em Inf. Comput.}, 202(2), 166--190, 2005.

\bibitem{ClementeEtAl2014}
Lorenzo Clemente, Fr{\'{e}}d{\'{e}}ric Herbreteau, and
            Gr{\'{e}}goire Sutre.
\newblock Decidable Topologies for Communicating Automata with {FIFO} and Bag Channels.
\newblock Proc. {\em 25th Int. Conf. on Concurrency Theory (CONCUR'14)},
\newblock Lecture Notes in Computer Science 8704, 281--296.
\newblock Springer, 2014.

\bibitem{DBLP:conf/apn/HaddadHM13}
  Serge Haddad,
               Rolf Hennicker, and
               Mikael H. M{\o}ller.
  \newblock Channel Properties of Asynchronously Composed {P}etri Nets.
  \newblock Proc. {\em Application and Theory of Petri Nets and Concurrency},
\newblock Lecture Notes in Computer Science 7927, 369--388.
\newblock Springer, 2013.

\bibitem{DBLP:conf/coordination/HennickerBD16}
Rolf Hennicker, Michel Bidoit, and Thanh{-}Son Dang.
\newblock On Synchronous and Asynchronous Compatibility of Communicating Components.
\newblock Proc. {\em 18th IFIP Int. Conf. on Coordination Models and Languages (COORDINATION'16)},
\newblock Lecture Notes in Computer Science 9686, 138--156.
\newblock Springer, 2016.

\bibitem{DBLP:conf/monterey/HennickerJK08}
  Rolf Hennicker,
               Stephan Janisch and
               Alexander Knapp.
  \newblock Refinement of Components in Connection-Safe Assemblies with Synchronous
               and Asynchronous Communication.
  \newblock {\em Foundations of Computer Software. Future Trends and Techniques for
               Development, 15th Monterey Workshop 2008},
\newblock Lecture Notes in Computer Science 6028, 154--180.
\newblock Springer, 2008.


\bibitem{DBLP:conf/tacas/TorreMP08}
  Salvatore {La Torre},
               P. Madhusudan, and
               Gennaro Parlato.
 \newblock Context-Bounded Analysis of Concurrent Queue Systems.
 \newblock Proc. {\em 14th Int. Conf. on Tools and Algorithms for the Construction and Analysis of Systems (TACAS'08)},
\newblock Lecture Notes in Computer Science 4963, 299--314.
\newblock Springer, 2008.

\bibitem{productlines}
  Kim Guldstrand Larsen,
               Ulrik Nyman, and
               Andrzej Wasowski.
  \newblock Modal {I/O} Automata for Interface and Product Line Theories.
  \newblock Proc.{\em 16th European Symposium on Programming,
               {ESOP'07}},
\newblock Lecture Notes in Computer Science 4421, 64--79.
\newblock Springer, 2007.


\bibitem{DBLP:conf/wsfm/LozesV11}
  {\'{E}}tienne Lozes and
               Jules Villard.
  \newblock Reliable Contracts for Unreliable Half-Duplex Communications.
  \newblock Proc. {\em 8th International Workshop on Web Services and Formal Methods {WS-FM}'11},
\newblock Lecture Notes in Computer Science 7176, 2--16.
\newblock Springer, 2011.


%\bibitem{DBLP:conf/stoc/Mayr81}
%  Ernst W. Mayr.
%  \newblock An Algorithm for the General {P}etri Net Reachability Problem.
%  \newblock Proc. {\em 13th Annual {ACM} Symposium on Theory of Computing},
%  \newblock 238--246.
%  \newblock {ACM}, 1981.

\bibitem{DBLP:journals/sosym/NorooziKMW15}
  Neda Noroozi,
               Ramtin Khosravi,
               Mohammad Reza Mousavi, and
               Tim A. C. Willemse.
  \newblock Synchrony and Asynchrony in Conformance Testing.
  \newblock  {\em Software and System Modeling}, 14 (1), 149-172, 2015.


\bibitem{DBLP:conf/facs2/OuederniSB13}
  Meriem Ouederni,
               Gwen Sala{\"{u}}n, and
               Tevfik Bultan.
  \newblock Compatibility Checking for Asynchronously Communicating Software.
  \newblock Proc. {\em Formal Aspects of Component Software - 10th International Symposium,
               {FACS}'13},
\newblock Lecture Notes in Computer Science 8348, 310--328.
\newblock Springer, 2013.

\bibitem{DBLP:journals/fuin/RacletBBCLP11}
  Jean{-}Baptiste Raclet,
               Eric Badouel,
               Albert Benveniste,
               Beno{\^{\i}}t Caillaud,
               Axel Legay, and
               Roberto Passerone.
  \newblock A Modal Interface Theory for Component-based Design.
  \newblock {\em Fundam. Inform.}, 108 (1-2), 119--149, 2011.

\bibitem{weiglhofer-diss}
  Martin Weiglhofer.
\newblock Automated Software Conformance Testing.
\newblock PhD thesis, TU Graz, 2009.

\bibitem{DBLP:journals/toplas/YellinS97}
   Daniel M. Yellin, and
   Robert E. Strom.
\newblock Protocol Specifications and Component Adaptors.
\newblock {\em {ACM} Trans. Program. Lang. Syst.}, 19 (2), 292-333, 1997.

 
\end{thebibliography}
\end{document}